\documentclass[12pt]{article}
\begin{document}
\baselineskip=0.7cm
\renewcommand{\theequation}{\arabic{section}.\arabic{equation}}
\renewcommand{\thesection}{\arabic{section}.}
\renewcommand{\thesubsection}{\arabic{section}.\arabic{subsection}}
\makeatletter
\def\section{\@startsection{section}{1}{\z@}{-3.5ex plus -1ex minus
 -.2ex}{2.3ex plus .2ex}{\large}}
\def\subsection{\@startsection{subsection}{2}{\z@}{-3.25ex plus -1ex minus
 -.2ex}{1.5ex plus .2ex}{\normalsize\it}}
\def\appendix{
\par
\setcounter{section}{0}
\setcounter{subsection}{0}
\def\thesection{\Alph{section}}}
\makeatother
\def\thefootnote{\fnsymbol{footnote}}
\begin{flushright}
hep-th/0002194\\
OU-HET 342, UT-877
\\
February 2000
\end{flushright}
\vspace{1cm}
\begin{center}
\Large
Constraints on effective Lagrangian of D-branes\\ from
non-commutative gauge theory

\vspace{1cm}
\normalsize
{\sc Yuji Okawa}${}^1$
\footnote{
E-mail:\ \ okawa@het.phys.sci.osaka-u.ac.jp}
and 
{\sc Seiji Terashima}${}^2$
\footnote{
E-mail:\ \ seiji@hep-th.phys.s.u-tokyo.ac.jp}
\\

\vspace{0.3cm}
{\it ${}^1$ Department of Physics,
Graduate School of Science, Osaka University\\
Toyonaka, Osaka 560-0043, Japan}

\vspace{0.3cm}
{\it ${}^2$ Department of Physics,
Faculty of Science, University of Tokyo\\
Tokyo 113-0033, Japan}

\vspace{1.3cm}
Abstract\\

\end{center}
It was argued that
there are two different descriptions of
the effective Lagrangian of gauge fields on D-branes
by non-commutative gauge theory
and by ordinary gauge theory
in the presence of a constant $B$ field background.
In the case of bosonic string theory, however,
it was found in the previous works that
the two descriptions are incompatible
under the field redefinition
which relates the non-commutative gauge field 
to the ordinary one found by Seiberg and Witten.
In this paper we resolve
this puzzle to observe the necessity
of gauge-invariant but $B$-dependent correction terms
involving metric in the field redefinition
which have not been considered before.
With the problem resolved, we establish
a systematic method under the $\alpha'$ expansion to derive
the constraints on the effective Lagrangian
imposed by the compatibility of the two descriptions
where the form of the field redefinition is not assumed.

\newpage
\section{Introduction}
\setcounter{equation}{0}
\indent

In the light of the recent developments in superstring and M-theory
brought by introducing D-branes,
it would be impossible to underestimate
the importance of understanding
the dynamics of collective coordinates of D-branes,
such as scalar fields on the worldvolume of the D-branes
representing their transverse positions and
gauge fields describing internal degrees of freedom.
In some situations or limits,
the effective Lagrangian describing such collective coordinates
is approximated or even supposed to be exactly described by
the dimensional reduction of super Yang-Mills theory from
ten dimensions to the worldvolume dimensions
of the D-branes \cite{bound-states, DKPS}.
For example, the matrix model of M-theory \cite{BFSS}
is based on the description of D0-branes in terms of
super Yang-Mills theory
and the most typical case
of the $AdS$/CFT correspondence \cite{Maldacena},
namely the correspondence between
$AdS_5$ and the four-dimensional super Yang-Mills theory,
is based on that of D3-branes.

Since the perturbative interactions between D-branes
and those between D-branes
and elementary excitations of strings
are completely defined
by the open string sigma model with the Dirichlet boundary condition,
it is in principle possible to calculate systematic corrections
of the effective Lagrangian
to the Yang-Mills theory.
For example, if we want to obtain the effective Lagrangian
of gauge fields on D-branes, we should calculate the S-matrix
of the scattering processes of the gauge fields
on D-branes in string theory,
then construct the effective Lagrangian such that it
reproduces the S-matrix correctly.
Another way to calculate the effective Lagrangian
is to calculate the beta function of the open string sigma model
with Dirichlet boundary condition
and to look for a Lagrangian whose equation of motion coincides
with the condition that the beta function vanishes.
The resulting Lagrangian is believed to coincide with
the one obtained from the string S-matrix
at least for tree-level processes.
However, the complexity of the calculation will necessarily increase
if we proceed to higher orders in the expansion with respect to
$\alpha'$ and the string coupling constant $g_s$
in the S-matrix approach
and to higher loops in the beta function approach
so that it would be helpful if other complementary approaches
to the effective Lagrangian are available.

Recently it is argued that
the effective Lagrangian of the gauge fields on D-branes
is described by non-commutative gauge theory
\cite{CDS}-\cite{Schomerus}
in the presence of a constant background field
of the Neveu-Schwarz--Neveu-Schwarz
two-form gauge field which is usually referred to as $B$ field.
It is also possible to describe it in terms of ordinary gauge theory,
however, the $B$-dependence in the two descriptions
is totally different
and it turned out that it is possible to constrain
the form of the effective Lagrangian
by the compatibility of the two descriptions.
Actually, it was shown in \cite{SW} that
the Dirac-Born-Infeld (DBI)
Lagrangian \cite{FT}-\cite{ACNY}
\footnote{
For a recent review of the Dirac-Born-Infeld theory
see \cite{Tseytlin} and references therein.}
satisfies the compatibility
in the approximation of neglecting derivatives of field strength
and its particular form was essential for the compatibility.
It is impossible to derive the DBI Lagrangian from
the gauge invariance alone so that
this shows that the requirement of the compatibility
does provide us with information on the dynamics of the gauge fields.

The proof of
the equivalence of the two descriptions
for the DBI Lagrangian in \cite{SW}
was beautiful,
however it is not clear how we can obtain the constraints
for other terms
imposed by the compatibility in a systematic way
so as to study how powerful and useful this approach will be.
This is our basic motivation of the present paper
and we will present a method to obtain the constraints systematically
in the $\alpha'$ expansion.

Actually a method towards this goal was developed
to some extent in \cite{Okawa}
where the problem of
whether it is possible to include
two-derivative corrections\footnote{
By $n$-derivative corrections to the DBI Lagrangian,
we mean terms with $n$ derivatives acting on
field strengths (not on gauge fields).
} 
to the DBI Lagrangian satisfying the compatibility
was discussed
and the most general form of the two-derivative corrections
up to the quartic order of field strength, $F^4$,
in the $\alpha'$ expansion was derived.
However there was a puzzle that the form of the two-derivative terms
which is consistent with the compatibility
disagreed with
the effective Lagrangian derived from bosonic string theory
although it was consistent with superstring theory.
Does this mean that the equivalence of the two descriptions
in the presence of a constant $B$ field
fails in bosonic string theory?
In the light of the argument in \cite{SW},
we do not think that it is the case.
It is most likely that we had made too strong assumptions
so that we only obtained a limited class of Lagrangians
which excludes that of bosonic string theory.
If it is the case, the methods which are currently available
such as the one in \cite{Okawa}
do not fulfill our purpose to derive the constraints correctly.
Furthermore the problem may not be limited to the case
of bosonic string theory.
Without resolving the puzzle, it would be dangerous
to develop discussions based on the equivalence
of the two descriptions.
Therefore we have to reconsider the assumptions which have been made
and find out the correct set of the assumptions from which
we should derive the constraints
using the problem of
whether the puzzle in bosonic strings is resolved
as a touchstone of the validity of our approach.

It will turn out that
the assumption which is not satisfied in bosonic strings
is the one on the form of the field redefinition
which relates the ordinary gauge field
to the non-commutative one.
The field redefinition
which preserves the gauge equivalence relation
found in \cite{SW} and further discussed in \cite{AK}
should be modified in general
and suffered from gauge-invariant
but $B$-dependent correction terms
involving metric.
In particular, our result will show that
such terms {\it must} exist in the case of bosonic string theory.
We will argue that the form of the field redefinition
should not be assumed as input when constraining
the form of the effective Lagrangian
and can be rather regarded as a consequence of the compatibility
of the two descriptions.
This argument is essential in resolving the puzzle
in the case of bosonic string theory as we will see.
We could jump into the problem of
two-derivative correction terms
to resolve the puzzle, however,
we will first determine the $F^4$ terms which 
coincide with those in the DBI Lagrangian correctly
without assuming the form of the field redefinition
in order to show that the idea presented in this paper is 
useful to constrain the effective Lagrangian.
We then apply it to two-derivative corrections
to resolve the puzzle
and the generalization to other cases would be straightforward.

The organization of this paper is as follows.
In Section 2, we first review the two descriptions of
the effective Lagrangian of the gauge fields on D-branes
in the presence of a constant $B$ field, namely,
the one in terms of ordinary gauge theory
and the one by non-commutative gauge theory,
to clarify what we assume when deriving the constraints.
We then derive the $F^4$ terms in the DBI Lagrangian
without assuming the form of the field redefinition
which relates the ordinary gauge field to the non-commutative one
in Section 3.
We extend our consideration to two-derivative corrections
to the DBI Lagrangian in Section 4
where the discrepancy in the case of bosonic string theory
is resolved by generalizing the form of the field redefinition.
Section 5 is devoted to conclusions and discussions.

\section{Review of the two descriptions in the presence of $B$}
\setcounter{equation}{0}
\indent

Let us first review the two descriptions
of the effective Lagrangian of D-branes
in the presence of a constant $B$ field background $B_{ij}$.
In this paper, we concentrate
on the effective Lagrangian of a gauge field on a single D-brane
in flat space-time, with constant metric $g_{ij}$, for simplicity.

The worldsheet action describing this system is
\begin{eqnarray}
S &=& \frac{1}{4 \pi \alpha'} \int_\Sigma d^2 \sigma
( g_{ij} \partial_a x^i \partial^a x^j
- 2 \pi i \alpha' B_{ij} \epsilon^{ab} \partial_a x^i \partial_b x^j )
\nonumber \\
&=& \frac{1}{4 \pi \alpha'} \int_\Sigma d^2 \sigma
g_{ij} \partial_a x^i \partial^a x^j
- \frac{i}{2} \int_{\partial \Sigma} d \tau
B_{ij} x^i \partial_\tau x^j,
\label{S}
\end{eqnarray}
where $\Sigma$ is the string worldsheet with Euclidean signature
and $\partial \Sigma$ is its boundary.
A background gauge field couples to the string worldsheet
by adding
\begin{equation}
S_{int} = -i \int_{\partial \Sigma} d \tau
A_i (x) \partial_\tau x^i
\label{S_int}
\end{equation}
to the action (\ref{S}).
Comparing (\ref{S}) and (\ref{S_int}),
we see that a constant $B$ field can be replaced by the gauge field
$$
A_i = -\frac{1}{2} B_{ij} x^j,
$$
whose field strength is $F_{ij} = B_{ij}$.
Thus we conclude that
there exists a definition of a gauge field in the effective Lagrangian
such that the effective Lagrangian depends on $B$ and $F$
only in the combination $B+F$ when we turn on a constant $B$ field.
This gauge field is an ordinary one, namely, the gauge transformations
and its field strength are defined by
\begin{eqnarray}
\delta_\lambda A_i &=& \partial_i \lambda, \\
F_{ij} &=& \partial_i A_j - \partial_j A_i, \\
\delta_\lambda F_{ij} &=& 0.
\end{eqnarray}
This is the first description of the effective Lagrangian
in terms of ordinary gauge theory.

To derive the second description in terms of
non-commutative gauge theory, let us examine
the propagator in (\ref{S}).
In the presence of a constant $B$ field,
the boundary condition of open strings is modified
and is no longer the Neumann one along the D-brane.
Thus the propagator in the sigma model is also modified
so as to satisfy the new boundary condition.
The explicit form of the propagator
evaluated at boundary points is \cite{FT}-\cite{ACNY}
\begin{equation}
\langle x^i (\tau) x^j (\tau') \rangle
= - \alpha' (G^{-1})^{ij} \log (\tau - \tau')^2
+ \frac{i}{2} \theta^{ij} \epsilon (\tau - \tau'),
\end{equation}
where the worldsheet is mapped to the upper half plane,
$\tau$ and $\tau'$ are points on the boundary and
\begin{eqnarray}
G_{ij} &=& g_{ij} - (2 \pi \alpha')^2 (B g^{-1} B)_{ij},
\\
\theta^{ij} &=&
- (2 \pi \alpha')^2 \left(
\frac{1}{g + 2 \pi \alpha' B} B \frac{1}{g - 2 \pi \alpha' B}
\right)^{ij}.
\end{eqnarray}
There are two important modifications here.
The first one is the coefficient in front of the log term
is no longer the metric $(g^{-1})^{ij}$.
The second one is the appearance of the term
proportional to the step function $\epsilon (\tau)$
which is $1$ or $-1$ for positive or negative $\tau$.

Now consider the $\theta$-dependence of 
correlation functions of open string vertex operators
which are given by
\begin{eqnarray}
&& \Big\langle
\prod_{n=1}^k
P_n ( \partial x (\tau_n), \partial^2 x (\tau_n), \ldots )
e^{i p^n \cdot x (\tau_n)}
\Big\rangle_{G,\theta} \nonumber \\
&& = \exp \Big(
-\frac{i}{2} \sum_{n > m} p^n_i \theta^{ij} p^m_j
\epsilon (\tau_n - \tau_m)
\Big) \nonumber \\
&& ~~~~~~~~ \times \Big\langle
\prod_{n=1}^k
P_n ( \partial x (\tau_n), \partial^2 x (\tau_n), \ldots )
e^{i p^n \cdot x (\tau_n)}
\Big\rangle_{G,\theta=0},
\label{theta-dependence}
\end{eqnarray}
where $P_n$'s are polynomials in derivatives of $x$
and $x$ are coordinates along the D-brane.
Since the second term in the propagator
does not contribute to contractions of derivatives of $x$,
the $\theta$-dependent part can be factorized
as the right-hand side of (\ref{theta-dependence}).
The string S-matrix can be obtained from these correlation
functions by putting external fields on shell
and integrating over the $\tau$'s.
Therefore, the S-matrix and the effective Lagrangian
constructed from it have a structure inherited from this form.

So we can see how the effective Lagrangian is modified
when we turn on the constant $B$ field.
To distinguish the gauge field
in this description from that in the preceding one,
let us rename it to $\hat{A}$
and denote the Lagrangian in terms of $\hat{A}$
as $\hat{{\cal L}}$.
The Lagrangian $\hat{{\cal L}}$ is constructed from
the one ${\cal L}$ in the absence of $B$ as follows.

First, the metric which appears when contracting
Lorentz indices is modified to $G_{ij}$ instead of $g_{ij}$
corresponding to the modification in the propagator.
Secondly, since the coupling constant can depend
on $B$, let us denote the coupling constant
in the presence of $B$ as $G_s$.
Finally, let us go on to the most important modification
related to the appearance of the $\theta$-dependent factor
\begin{equation}
\exp \Big(
-\frac{i}{2} \sum_{n > m} p^n_i \theta^{ij} p^m_j
\epsilon (\tau_n - \tau_m)
\Big)
\end{equation}
in (\ref{theta-dependence}).
It corresponds to modifying the ordinary product of functions
to the associative but non-commutative $\ast$ product defined by
\begin{equation}
f(x) \ast g(x) = \exp \left. \left(
\frac{i}{2} \theta^{ij}
\frac{\partial}{\partial \xi^i}
\frac{\partial}{\partial \zeta^j} \right)
f(x + \xi) g(x + \zeta) \right|_{\xi=\zeta=0},
\end{equation}
in the momentum-space representation.
Now the $B$-dependence of the effective Lagrangian
in this description can be obtained
through the following replacements:
$A$ by $\hat{A}$, $g_{ij}$ by $G_{ij}$, $g_s$ by $G_s$
and ordinary multiplication by the $\ast$ product.
Corresponding to the modification of the product,
the gauge transformations and the definition of field strength
are also modified as follows:
\begin{eqnarray}
\hat{\delta}_{\hat{\lambda}} \hat{A}_i
&=& \partial_i \hat{\lambda} + i \hat{\lambda} \ast \hat{A}_i
- i \hat{A}_i \ast \hat{\lambda},
\label{non-commutative-transformation} \\
\hat{F}_{ij} &=& \partial_i \hat{A}_j - \partial_j \hat{A}_i
-i \hat{A}_i \ast \hat{A}_j +i \hat{A}_j \ast \hat{A}_i,
\label{definition-F-hat} \\
\hat{\delta}_{\hat{\lambda}} \hat{F}_{ij}
&=& i \hat{\lambda} \ast \hat{F}_{ij}
- i \hat{F}_{ij} \ast \hat{\lambda}.
\end{eqnarray}

We have seen that there are two different
effective Lagrangians of the gauge field on the D-brane
which reproduce the S-matrix of string theory
in the presence of a constant $B$.
What we have learned from the action (\ref{S})
and the interaction (\ref{S_int})
can be summarized as follows.
\begin{enumerate}
\item
There exists a definition of a gauge field $A_i$ such that
the Lagrangian in terms of it respects
the ordinary gauge invariance
and it depends on $B$ only in the combination $B+F$.
\item
There exists a definition of a gauge field $\hat{A}_i$ such that
the Lagrangian in terms of it respects
the non-commutative gauge invariance
and it depends on $B$ only through
$G_{ij}$, $G_s$ and $\theta^{ij}$
in the non-commutative $\ast$ product.
\begin{equation}
\label{two}
\end{equation}
\end{enumerate}
These are our fundamental assumptions
and we will consider constraints on the form of
the effective Lagrangian imposed by the compatibility of
them in what follows.

It is not surprising that there are different descriptions
of the effective Lagrangian
since the S-matrix is unchanged under field redefinitions
in the effective Lagrangian so that
the construction of the effective Lagrangian from
the S-matrix elements is always subject to an ambiguity
originated in the field redefinitions.
Thus we do not expect that the two gauge fields $A_i$ and $\hat{A}_i$
coincide: they would be related by a field redefinition.
Usually we consider field redefinitions of the form
$$
A_i \rightarrow A_i + f_i (\partial,F),
$$
where $f_i (\partial,F)$ denotes an arbitrary gauge-invariant
expression made of $F_{ij}$, $\partial_k F_{ij}$,
$\partial_k \partial_l F_{ij}$, and so on.
The field redefinitions of this kind
preserve the ordinary gauge invariance.
However they will not work in this case because
the gauge transformation
of $\hat{A}_i$ is different from that of $A_i$.
The field redefinition which relates $\hat{A}_i$ to $A_i$
must preserve the gauge equivalence relation,
namely it satisfies
\begin{equation}
\hat{A} (A) + \hat{\delta}_{\hat{\lambda}} \hat{A} (A)
= \hat{A} ( A + \delta_\lambda A ),
\label{A-hat}
\end{equation}
with infinitesimal $\lambda$ and $\hat{\lambda}$.
Whether there exists a field redefinition which satisfies
(\ref{A-hat}) is a nontrivial question,
however, a perturbative solution with respect to $\theta$
was found by Seiberg and Witten \cite{SW}.
Its explicit form for the rank-one case
is given by\footnote{
Solutions to the gauge equivalence relation
were further discussed in \cite{AK}.
}
\begin{eqnarray}
\hat{A}_i &=& A_i - \frac{1}{2} \theta^{kl}
A_k (\partial_l A_i + F_{li}) + O(\theta^2),
\label{A-hat-2} \\
\hat{\lambda} &=& \lambda + \frac{1}{2} \theta^{kl}
\partial_k \lambda A_l + O(\theta^2).
\end{eqnarray}
However we should emphasize here that
{\em we do not assume the explicit form of
the field redefinition which relates $\hat{A}_i$ to $A_i$
when we derive constraints on the form of the effective Lagrangian
in the present paper.}
What we assume is the two assumptions (\ref{two}) alone.
This is an important difference from the previous works
such as \cite{SW} or \cite{Okawa}.
The form of the field redefinition is
rather regarded as a consequence of the compatibility
of the two descriptions in terms of ordinary and non-commutative
gauge theories as we will see in the next section.

Before proceeding,
we should make a comment on the relation between
our assumptions (\ref{two}) and regularization schemes
in the sigma model.
We mentioned the ambiguity related to field redefinitions
in constructing effective Lagrangian from S-matrix elements.
In the case of string theory,
we can also understand the origin of the ambiguity
in the point of view of the sigma model
to be coming from degrees of freedom to choose
different regularization schemes
as was discussed in \cite{SW}.
We arrived at the assumptions (\ref{two}) from the properties
of (\ref{S}) and (\ref{S_int}) at classical level.
However it is necessary to regularize the theory
to define composite operators
such as (\ref{S_int}) at quantum level.
The description in terms of the ordinary gauge field $A_i$
will be derived from a Pauli-Villars type regularization
while the description in terms of the non-commutative gauge
field $\hat{A}_i$ will be derived from a point-splitting type
regularization.
However if we take the simple point-splitting regularization
discussed in \cite{SW}
in which we cut out the region $| \tau - \tau' | < \delta$
and take the limit $\delta \to 0$,
the non-commutative gauge transformation suffers from
$\alpha'$ corrections before taking the zero slope limit.
Therefore it is not clear whether there is
an appropriate regularization corresponding
to the non-commutative gauge field $\hat{A}_i$
in the second assumption of (\ref{two})
where no zero slope limit is taken.
In this sense, we regard (\ref{two}) as assumptions
although we can argue that they are plausible
in the following way.
If the effective action before turning on $B$
is invariant under the ordinary gauge transformation
and the $B$-dependence can be made only through
$G_{ij}$, $G_s$ and $\theta^{ij}$,
the action after turning on $B$ is automatically invariant
under the non-commutative gauge transformation
(\ref{non-commutative-transformation})
at least for the case where
the rank of the gauge group is greater than one.
The case with the rank-one gauge theory may be slightly subtle
but it would be naturally expected that
it holds in this case as well.
At any rate, our basic standpoint is that
the effective Lagrangian we discuss in this paper is
constructed so as to reproduce the S-matrix elements correctly
and it is not necessary to consider its relation
to the background field in the sigma model in what follows.

\section{Determination of $F^4$ terms revisited}
\setcounter{equation}{0}

\subsection{Determination without assuming
the form of the field redefinition}
\indent

Let us now proceed to see how the $F^4$ terms in the DBI Lagrangian
are determined by the assumptions (\ref{two})
although the form of the field redefinition is not assumed.
Since we study the effective Lagrangian
in the $\alpha'$ expansion,
we present the following formulas for convenience
which will be used repeatedly:
\begin{eqnarray}
(G^{-1})^{ij} &=& (g^{-1})^{ij}
+ (2 \pi \alpha')^2
( g^{-1} B g^{-1} B g^{-1} )^{ij}
+ O(\alpha'^4),
\label{G-expansion}
\\
\theta^{ij} &=& - (2 \pi \alpha')^2
( g^{-1} B g^{-1} )^{ij} + O(\alpha'^4), \\
f \ast g &=& fg -\frac{i}{2} (2 \pi \alpha')^2
( g^{-1} B g^{-1} )^{kl} \partial_k f \partial_l g
+ O(\alpha'^4).
\label{star-expansion}
\end{eqnarray}
The lowest order term of the effective Lagrangian of a gauge field
on a D-brane in the $\alpha'$ expansion is the $F^2$ term:
\begin{eqnarray}
{\cal L}(F) &=& \frac{\sqrt{\det g}}{g_s} \left[
(g^{-1})^{ij} F_{jk} (g^{-1})^{kl} F_{li}
+ O(\alpha') \right] \nonumber \\
&=& \frac{\sqrt{\det g}}{g_s} \left[
F_{ij} F_{ji}
+ O(\alpha') \right] \nonumber \\
&\equiv& \frac{\sqrt{\det g}}{g_s} \left[
{\rm Tr} F^2
+ O(\alpha') \right].
\label{F2}
\end{eqnarray}
Here we omitted a possible overall factor
including an appropriate power of $\alpha'$.
Since the discussions presented in this paper
do not depend on the dimension of space-time
on which the gauge theory is defined,
namely, the dimension of worldvolume of the D-brane,
if we want to supply the overall factor,
we only need to multiply an appropriate power of
$\alpha'$ to the Lagrangian to make the action
dimensionless and a numerical constant
which depends on the convention.
In the second line of (\ref{F2}),
we made $g^{-1}$ implicit as
\begin{equation}
A_i B_i \equiv (g^{-1})^{ij} A_i B_j.
\label{implicit-g-1}
\end{equation}
Since Lorentz indices in most of the expressions in what follows
are contracted with respect to the metric $g_{ij}$,
we will adopt this convention together with
\begin{equation}
\partial^2 \equiv (g^{-1})^{ij} \partial_i \partial_j,
\label{implicit-g-2}
\end{equation}
to simplify the expressions
unless the other metric $G_{ij}$ is explicitly used.
And Tr denotes the trace over Lorentz indices
as can be seen from the third line of (\ref{F2}).

Now the assumptions (\ref{two}) imply that
we can describe the system
in two different ways when we turn on $B$
as follows:
\begin{eqnarray}
{\cal L} (B+F)
&=& \frac{\sqrt{\det g}}{g_s} \left[
{\rm Tr} (B+F)^2
+ O(\alpha') \right],
\label{(B+F)^2} \\
\hat{{\cal L}} (\hat{F})
&=& \frac{\sqrt{\det G}}{G_s} \left[
(G^{-1})^{ij} \hat{F}_{jk} \ast (G^{-1})^{kl} \hat{F}_{li}
+ O(\alpha') \right].
\label{F-hat^2}
\end{eqnarray}
In the case of higher-rank gauge theory, it follows from
the comparison between (\ref{(B+F)^2}) and (\ref{F-hat^2})
when $B$ vanishes that
\begin{eqnarray}
G_s &=& g_s + O(\alpha'),
\label{G_s-lowest} \\
\hat{A}_i &=& A_i + O(\alpha').
\label{A-hat-lowest}
\end{eqnarray}
In the rank-one case, on the other hand, we can only determine
the normalizations of $G_s$ and $\hat{A}_i$ as
\begin{eqnarray}
G_s &=& t g_s + O(\alpha'),
\label{G_s-0}
\\
\hat{A}_i &=&  \sqrt{t} A_i + O(\alpha'),
\label{U1}
\end{eqnarray}
        from the consideration at the lowest order
in $\alpha'$ alone
since there is no interaction in the $F^2$ term.
The normalizations of $A_i$ and $\hat{A}_i$
and hence that of $G_s$ are already determined by (\ref{two})
since if we rescale $A_i$ or $\hat{A}_i$ then the $B$-dependence
does not take the combination $B+F$ for the description
in terms of $A_i$
and the field strength $\hat{F}_{ij}$ does not take the form
(\ref{definition-F-hat}) anymore
as for the description using $\hat{A}_i$.
Therefore we can in principle determine the constant $t$ from
the assumptions (\ref{two}).
However the calculation for the determination of $t$ is
slightly messy so that
we will defer it to Appendix A
and proceed assuming $t=1$ in this section
for the sake of brevity
which will be justified in Appendix A.

Let us first check that ${\cal L} (B+F)$ and $\hat{{\cal L}} (\hat{F})$
coincide at the lowest order in $\alpha'$,
which is necessary to be consistent with (\ref{two}).
In general,
the Lagrangian $\hat{{\cal L}}$ on the non-commutative side
reduces to the one ${\cal L}$ on the commutative side
at the lowest order in $\alpha'$.
In this case,
\begin{equation}
(G^{-1})^{ij} \hat{F}_{jk} \ast (G^{-1})^{kl} \hat{F}_{li}
= ( \partial_i \hat{A}_j  - \partial_j \hat{A}_i )
  ( \partial_j \hat{A}_i  - \partial_i \hat{A}_j )
  + O(\alpha'^2).
\end{equation}
What is less trivial is the question whether
${\rm Tr} (B+F)^2$ reduces to ${\rm Tr} F^2$
up to total derivative, namely,
whether ${\rm Tr} F^2$ satisfies the {\it initial term condition}
defined by
\begin{equation}
f(B+F) = f(F) + {\rm total~derivative},
\label{initial}
\end{equation}
in \cite{Okawa}, which is the condition for a term
to be qualified as an initial term of a consistent Lagrangian
in the $\alpha'$ expansion.
It is verified that ${\rm Tr} F^2$ satisfies this condition
as follows:
\begin{eqnarray}
&& {\rm Tr} (B+F)^2
\nonumber \\
&=& {\rm Tr} F^2
+2 {\rm Tr} BF
+ {\rm Tr} B^2
\nonumber \\
&=& {\rm Tr} F^2 
+ {\rm total~derivative} + {\rm const}.
\end{eqnarray}

The $F^4$ terms in the DBI Lagrangian
are determined by the consideration
at the next order terms in the $\alpha'$ expansion of (\ref{F-hat^2}),
which are given by
\begin{eqnarray}
&& \frac{\sqrt{\det G}}{G_s}
(G^{-1})^{ij} \hat{F}_{jk} \ast (G^{-1})^{kl} \hat{F}_{li}
\nonumber \\
&=& \frac{\sqrt{\det g}}{G_s} \Biggl[
( \partial_i \hat{A}_j  - \partial_j \hat{A}_i )
( \partial_j \hat{A}_i  - \partial_i \hat{A}_j )
-4 (2 \pi \alpha')^2 B_{kl}
\partial_k \hat{A}_i \partial_l \hat{A}_j \partial_j \hat{A}_i
\nonumber \\
&& +2 (2 \pi \alpha')^2 (B^2)_{ij}
( \partial_j \hat{A}_k  - \partial_k \hat{A}_j )
( \partial_k \hat{A}_i  - \partial_i \hat{A}_k )
\nonumber \\
&& -\frac{1}{2} (2 \pi \alpha')^2 {\rm Tr} B^2
( \partial_i \hat{A}_j  - \partial_j \hat{A}_i )
( \partial_j \hat{A}_i  - \partial_i \hat{A}_j )
+ O(\alpha'^4) \Biggr].
\label{F-hat^2-2}
\end{eqnarray}
What is important here is the existence of the second term
on the right-hand side of (\ref{F-hat^2-2}).
It gives a non-vanishing contribution to the three-point scattering
amplitude of the gauge fields.
More precisely, if we represent the asymptotic fields
in $N$-point scattering as
\begin{eqnarray}
A^{{\rm asym}~a}_i (x) = \zeta^a_i e^{i k^a \cdot x},
\qquad a= 1, 2, \ldots, N,
\label{configuration} \\
( k^a )^2 =0, \quad \zeta^a \cdot k^a =0, \quad
\sum_{a=1}^N k^a_i =0,
\label{on-shell-condition}
\end{eqnarray}
the second term on the right-hand side
of (\ref{F-hat^2-2}) gives a contribution
to the three-point amplitude
of order $O(B, \zeta^3, k^3)$.
It can be easily shown that
no other term can produce the contribution of this form
on the non-commutative side.
Therefore this contribution cannot be canceled
and must be reproduced from the Lagrangian ${\cal L} (B+F)$
on the commutative side.

There are two terms on the commutative side
which can produce the $O(B, \zeta^3, k^3)$ contribution
to the three-point amplitude.
They are ${\rm Tr} (B+F)^4$ and $[{\rm Tr} (B+F)^2]^2$:
\begin{eqnarray}
{\rm Tr} (B+F)^4 &=& {\rm Tr} F^4
+ 4 {\rm Tr} BF^3 + O(B^2), \\
\left[{\rm Tr} (B+F)^2 \right]^2
&=& ( {\rm Tr} F^2 )^2
+ 4 {\rm Tr} BF {\rm Tr} F^2 + O(B^2).
\end{eqnarray}
There are several terms in ${\rm Tr} BF^3$
and ${\rm Tr} BF {\rm Tr} F^2$
when we expand them as $F_{ij} = \partial_i A_j - \partial_j A_i$,
but some of them
which contain $\partial^2 A_i$
or $\partial_i A_i$ do not contribute to the S-matrix
of the three-point scattering
because of the on-shell conditions
$k^2 =0$ and $\zeta \cdot k =0$.
Moreover, it will be useful to observe that
terms of the form $f \partial_i g \partial_i h$ in general
do not contribute to the S-matrix of three-point scattering
where $f$, $g$ and $h$ are massless fields or their derivatives.
This follows from the fact
$k^1 \cdot k^2 = k^2 \cdot k^3 = k^3 \cdot k^1 =0$
which can be easily seen as
\begin{equation}
0 = ( k^3 )^2 = ( k^1 + k^2 )^2 = 2 k^1 \cdot k^2,
\end{equation}
where we used $k^1 + k^2 + k^3 =0$ and $( k^a )^2 =0$.
Another way to see this is to rewrite $f \partial_i g \partial_i h$
as follows:
\begin{equation}
f \partial_i g \partial_i h
= \frac{1}{2}
(\partial^2 f g h - f \partial^2 g h - f g \partial^2 h)
+ \frac{1}{2} \partial^2 (f g h) - \partial_i ( \partial_i f  g h).
\label{formula}
\end{equation}
Having been equipped with this formula,
we can extract the part which contributes to the S-matrix from
${\rm Tr} BF^3$ and ${\rm Tr} BF {\rm Tr} F^2$ as follows:
\begin{eqnarray}
{\rm Tr} BF^3 &=& B_{ij} F_{jk} F_{kl} F_{li}
\nonumber \\
&=& 2 B_{ij} \partial_j A_k \partial_k A_l \partial_l A_i
-2 B_{ij} \partial_j A_k \partial_k A_l \partial_i A_l
\nonumber \\
&& + {\rm ~terms~with~} \partial^2 A
+ {\rm total~derivative},
\\
{\rm Tr} BF {\rm Tr} F^2 &=& B_{ij} F_{ji} F_{kl} F_{lk}
\nonumber \\
&=& 4 B_{ij} \partial_j A_i \partial_k A_l \partial_l A_k
\nonumber \\
&=& -8 B_{ij} A_i \partial_k A_l \partial_j \partial_l A_k
+ {\rm total~derivative}
\nonumber \\
&=& 8 B_{ij} \partial_l A_i \partial_k A_l \partial_j A_k
+ {\rm a~term~with~} \partial_l A_l
+ {\rm total~derivative}.
\end{eqnarray}
To summarize, we have found that all the terms which contribute
to the S-matrix of order $O(B, \zeta^3, k^3)$.
On the non-commutative side, there was only one source,
$$
{\rm Tr} ( G^{-1} \hat{F} \ast G^{-1} \hat{F} ) \to
-4 (2 \pi \alpha')^2 B_{kl}
\partial_k \hat{A}_i \partial_l \hat{A}_j \partial_j \hat{A}_i,
$$
while there were two on the commutative side:
\begin{eqnarray*}
{\rm Tr} BF^3 &\to&
 2 B_{ij} \partial_j A_k \partial_k A_l \partial_l A_i
-2 B_{ij} \partial_i A_l \partial_j A_k \partial_k A_l,
\\
{\rm Tr} BF {\rm Tr} F^2 &\to&
8 B_{ij} \partial_j A_k \partial_k A_l \partial_l A_i.
\end{eqnarray*}
It is not difficult to show that
the contributions to the S-matrix from
$B_{ij} \partial_j A_k \partial_k A_l \partial_l A_i$
and $B_{ij} \partial_i A_l \partial_j A_k \partial_k A_l$
are non-vanishing and linearly independent.
Thus the conclusion derived from (\ref{two}) is that
to reproduce the contribution to the S-matrix from
the Lagrangian $\hat{{\cal L}} (\hat{F})$,
the following terms must exist
in the Lagrangian ${\cal L} (B+F)$:
\begin{equation}
2 (2 \pi \alpha')^2 {\rm Tr} BF^3
-\frac{1}{2} (2 \pi \alpha')^2 {\rm Tr} BF {\rm Tr} F^2.
\label{BF^3}
\end{equation}
We can uniquely construct the Lagrangian ${\cal L} (F)$
such that ${\cal L} (B+F)$ generates the terms (\ref{BF^3}),
which is given by
\begin{eqnarray}
{\cal L}(F) = \frac{\sqrt{\det g}}{g_s} \Biggl[
{\rm Tr} F^2
+ (2 \pi \alpha')^2 \left[
\frac{1}{2} {\rm Tr} F^4
-\frac{1}{8} ({\rm Tr} F^2)^2 \right]
\nonumber \\
+ ~O(\alpha'^4) + {\rm derivative~corrections~}
\Biggr].
\end{eqnarray}
This coincides with the $\alpha'$ expansion of
the DBI Lagrangian for a single D$p$-brane,
\begin{equation}
{\cal L}_{DBI} (F) =
\frac{1}{g_s (2 \pi)^p (\alpha')^{(p+1)/2}}
\sqrt{\det (g + 2 \pi \alpha' F)},
\label{DBI}
\end{equation}
up to an overall factor and an additive constant.
Thus we have succeeded in determining the $F^4$ terms
in the DBI Lagrangian from the assumptions (\ref{two})
without referring to the explicit form of the field redefinition
which relates $\hat{A}_i$ to $A_i$.
We will derive its form in the next subsection.

\subsection{Field redefinition}
\indent

We have seen that the two effective Lagrangians,
\begin{eqnarray}
{\cal L}(B+F) = \frac{\sqrt{\det g}}{g_s} \Biggl[
{\rm Tr} (B+F)^2
+ (2 \pi \alpha')^2 \left[
\frac{1}{2} {\rm Tr} (B+F)^4
-\frac{1}{8} [ {\rm Tr} (B+F)^2 ]^2 \right]
\nonumber \\
+ ~O(\alpha'^4) + {\rm derivative~corrections~}
\Biggr],
\end{eqnarray}
and
\begin{eqnarray}
\hat{{\cal L}}(\hat{F}) = \frac{\sqrt{\det G}}{G_s} \Biggl[
{\rm Tr} ( G^{-1} \hat{F} \ast G^{-1} \hat{F} )
+ (2 \pi \alpha')^2 \biggl[
\frac{1}{2} {\rm Tr} ( G^{-1} \hat{F} )^4_{\rm arbitrary}
\nonumber \\
-\frac{1}{8} ( {\rm Tr} ( G^{-1} \hat{F} )^2 )^2_{\rm arbitrary}
\biggr]
+ ~O(\alpha'^4) + {\rm derivative~corrections~}
\Biggr],
\label{L-hat-4}
\end{eqnarray}
produce the same contribution to the S-matrix
of order $O(B, \zeta^3, k^3)$.
Here we added the $\hat{F}^4$ terms to $\hat{{\cal L}}(\hat{F})$
which were required by the existence of the corresponding
$F^4$ terms in ${\cal L}(B+F)$
and the subscripts ``arbitrary'' there mean
that the ordering of the four field strengths
in each term is arbitrary.
Since the $\ast$ product is non-commutative,
we have to specify the ordering of field strengths
as in the case of the ordinary Yang-Mills theory.
However, there is no principle in determining the ordering
for the rank-one case and we leave it arbitrary for now.
The fact that two effective Lagrangians produce
the same contribution to the S-matrix
does not mean that the two must coincide at off-shell level
but implies that the fields in the two effective Lagrangians
can be related by a field redefinition.
Let us see this explicitly for the case in hand.
By expanding the $O(\alpha'^2)$ terms in ${\cal L}(B+F)$,
we have
\begin{eqnarray}
&& \frac{1}{2} (2 \pi \alpha')^2 {\rm Tr} (B+F)^4
-\frac{1}{8} (2 \pi \alpha')^2 [ {\rm Tr} (B+F)^2 ]^2
\nonumber \\
&=& \frac{1}{2} (2 \pi \alpha')^2 {\rm Tr} F^4
-\frac{1}{8} (2 \pi \alpha')^2 ( {\rm Tr} F^2 )^2
\nonumber \\
&& +2 (2 \pi \alpha')^2 {\rm Tr} BF^3
-\frac{1}{2} (2 \pi \alpha')^2 {\rm Tr} BF {\rm Tr} F^2
\nonumber \\
&& +2 (2 \pi \alpha')^2 {\rm Tr} B^2 F^2
-\frac{1}{4} (2 \pi \alpha')^2 {\rm Tr} B^2 {\rm Tr} F^2
+ {\rm total~derivative} + {\rm const.},
\label{(B+F)^4}
\end{eqnarray}
where we used the fact that
\begin{equation}
(2 \pi \alpha')^2 \left[
{\rm Tr} ( B F )^2
- \frac{1}{2} ( {\rm Tr} B F )^2
\right] = {\rm total~derivative}.
\end{equation}
Obviously the $O(B)$ and $O(B^2)$ parts of (\ref{(B+F)^4})
do not coincide with those of (\ref{F-hat^2-2})
if we assume $\hat{A}_i=A_i$.
Let us first consider
the difference in the $O(B)$ part:
\begin{eqnarray}
\Delta {\cal L} &\equiv&
2 (2 \pi \alpha')^2 {\rm Tr} BF^3
-\frac{1}{2} (2 \pi \alpha')^2 {\rm Tr} BF {\rm Tr} F^2
- \Bigl( -4 (2 \pi \alpha')^2 B_{kl}
\partial_k A_i \partial_l A_j \partial_j A_i \Bigr)
\nonumber \\
&=& 2 (2 \pi \alpha')^2 B_{kl} F_{lj} F_{ji} F_{ik}
+2 (2 \pi \alpha')^2 B_{kl} A_k \partial_l F_{ij} F_{ji}
+2 (2 \pi \alpha')^2 B_{kl} \partial_k A_i \partial_l A_j F_{ji}
\nonumber \\
&& + {\rm ~total~derivative}
\nonumber \\
&=& 2 (2 \pi \alpha')^2 B_{kl} F_{ji}
( F_{lj} F_{ik} + A_k \partial_l F_{ij}
+ \partial_k A_i \partial_l A_j )
+ {\rm total~derivative}.
\end{eqnarray}
This must be reduced to the field redefinition
which relates $\hat{A}_i$ to $A_i$.
We can make it manifest by noting the fact that
\begin{eqnarray}
&& B_{kl} F_{ji} \partial_i [
A_k (\partial_l A_j + F_{lj})]
\nonumber \\
&=& B_{kl} F_{ji} [
\partial_i A_k (\partial_l A_j + F_{lj})
+ A_k (\partial_l \partial_i A_j + \partial_i F_{lj})]
\nonumber \\
&=& B_{kl} F_{ji} \left[
(F_{ik} + \partial_k A_i) (F_{lj} + \partial_l A_j)
+ A_k \left( \frac{1}{2} \partial_l F_{ij} + \partial_i F_{lj}
\right) \right]
\nonumber \\
&=& B_{kl} F_{ji} ( F_{lj} F_{ik} + A_k \partial_l F_{ij}
+ \partial_k A_i \partial_l A_j ),
\label{transformation}
\end{eqnarray}
where we used the facts that
\begin{equation}
F_{ji} \partial_i F_{lj} = \frac{1}{2} F_{ji} \partial_l F_{ij},
\end{equation}
and that
\begin{equation}
B_{kl} F_{ji} (F_{ik} \partial_l A_j + \partial_k A_i F_{lj}) =0.
\end{equation}
Then the difference $\Delta {\cal L}$ can be rewritten
using (\ref{transformation}) as
\begin{eqnarray}
\Delta {\cal L} &=&
2 (2 \pi \alpha')^2 B_{kl} F_{ji}
\partial_i [ A_k (\partial_l A_j + F_{lj})]
+ {\rm total~derivative}
\nonumber \\
&=& 2 (2 \pi \alpha')^2 B_{kl} \partial_i F_{ij}
A_k (\partial_l A_j + F_{lj})
+ {\rm total~derivative}.
\end{eqnarray}
The fact that $\Delta {\cal L}$ does not contribute
to the S-matrix and can be reduced to the field redefinition
of $\hat{A}_i$ is now manifest in this form
since $\Delta {\cal L}$ is proportional to $\partial_i F_{ij}$
and hence vanishes using the equation of motion.
If we write
\begin{equation}
\hat{A}_i = A_i + (2 \pi \alpha')^2 \Delta A_i + O(\alpha'^4),
\end{equation}
it obeys that
\begin{equation}
(\partial_i \hat{A}_j - \partial_j \hat{A}_i)
(\partial_j \hat{A}_i - \partial_i \hat{A}_j)
= F_{ij} F_{ji}
+ 4 (2 \pi \alpha')^2 \partial_i F_{ij} \Delta A_j
+ O(\alpha'^4).
\end{equation}
Thus the appropriate field redefinition is determined by
solving the equation
\begin{equation}
4 (2 \pi \alpha')^2 \partial_i F_{ij} \Delta A_j
= \Delta {\cal L}.
\end{equation}
The solution is given by
\begin{equation}
\Delta A_i = \frac{1}{2} B_{kl} A_k (\partial_l A_i + F_{li}),
\end{equation}
up to gauge transformations,
and the relation between $\hat{A}_i$ and $A_i$ is
\begin{equation}
\hat{A}_i = A_i + \frac{1}{2} (2 \pi \alpha')^2
B_{kl} A_k (\partial_l A_i + F_{li})
+ O(\alpha'^4).
\label{redefinition-1}
\end{equation}
This precisely coincides with the field redefinition
(\ref{A-hat-2}) found by Seiberg and Witten \cite{SW}
if we express $\theta$ in terms of $B$.
This was expected
since we assumed in (\ref{two})
the ordinary gauge invariance
in the description in terms of $A_i$ and
the non-commutative gauge invariance
in the description using $\hat{A}_i$
so that the gauge equivalence relation (\ref{A-hat})
must be satisfied.
Our result is therefore consistent with the previous works.
However it is important to note that
this form of the field redefinition should be regarded as
a consequence of the assumptions (\ref{two}) in our approach.
We did not have to know the form of the field redefinition
in the determination of the $F^4$ terms
and the form of the field redefinition was determined from
the difference between the two effective Lagrangians
at off-shell level.

The $O(B^2)$ part of the difference
between (\ref{F-hat^2-2}) and (\ref{(B+F)^4}),
\begin{equation}
\frac{1}{4} (2 \pi \alpha')^2 {\rm Tr} B^2 {\rm Tr} F^2,
\label{superfluous-B}
\end{equation}
is proportional to the $F^2$ term so that
it can be absorbed into the definition of $G_s$ as follows:
\begin{equation}
G_s
= g_s \left[
1 -\frac{1}{4} (2 \pi \alpha')^2
{\rm Tr} B^2 + O(\alpha'^4)
\right].
\label{G_s}
\end{equation}
Here it is also possible to take care of the difference
(\ref{superfluous-B}) by a field redefinition of $\hat{A}_i$
just as in the case of the difference in the $O(B)$ part
and we cannot determine how we should treat
(\ref{superfluous-B}) from the consideration
at the order $\alpha'^2$.
However since the normalizations of $A_i$ and $\hat{A}_i$
are already determined by (\ref{two}) as we mentioned
below (\ref{U1}), the ambiguity must be fixed
by the consideration at higher orders.
We will determine the $O(\alpha'^2)$ part of $G_s$
in Appendix B from the consideration at order $\alpha'^4$,
which justifies (\ref{G_s}).

We have demonstrated how to constrain the effective Lagrangian
of gauge fields on D-branes from the assumptions (\ref{two})
for the $F^4$ terms in the DBI Lagrangian.
We should now proceed to the reconsideration
of the constraints on the two-derivative
corrections to the DBI Lagrangian
where the discrepancy was found
in the case of bosonic string theory \cite{Okawa}.

\section{Constraints on two-derivative corrections}
\setcounter{equation}{0}

\subsection{$O(\alpha')$ terms}
\indent

The two-derivative corrections to the DBI Lagrangian
can first appear at order $\alpha'$
compared with the $F^2$ term.
Let us first survey possible terms at this order
in both ordinary and non-commutative gauge theories.

In ordinary gauge theory, Lagrangians are made of
field strength and its derivatives.
At order $\alpha'$,
terms of the forms $\partial F \partial F$,
$F \partial^2 F$ and $F^3$ are possible.
However since the $F \partial^2 F$ terms can be transformed
to the $\partial F \partial F$ terms
using the integration by parts
and $F^3$ terms vanish for the rank-one case,
it is sufficient to consider the $\partial F \partial F$ terms.
There are three different ways to contract Lorentz indices:
\begin{eqnarray}
T_1 \equiv \partial_i F_{ik} \partial_j F_{jk}, \qquad
T_2 \equiv \partial_j F_{ik} \partial_i F_{jk}, \qquad
T_3 \equiv \partial_k F_{ij} \partial_k F_{ji}.
\end{eqnarray}
Using the Bianchi identity, the term $T_3$ reduces to $T_2$,
\begin{equation}
T_3 = -2 T_2,
\end{equation}
and the two remaining terms $T_1$ and $T_2$
coincide up to total derivative:
\begin{eqnarray}
T_1 &=& - F_{ik} \partial_i \partial_j F_{jk}
+ {\rm total~derivative}, \\
T_2 &=& - F_{ik} \partial_j \partial_i F_{jk}
+ {\rm total~derivative}.
\end{eqnarray}
Thus any term at order $\alpha'$
can be transformed to $T_1$.

The story is slightly different in non-commutative gauge theory.
The building blocks of Lagrangians in non-commutative gauge theory
are field strength $\hat{F}$
and its covariant derivatives defined by
\begin{equation}
\hat{D}_i \hat{F}_{jk} =
\partial_i \hat{F}_{jk} -i \hat{A}_i \ast \hat{F}_{jk}
+i \hat{F}_{jk} \ast \hat{A}_i.
\end{equation}
At order $\alpha'$,
terms of the forms $\hat{D} \hat{F} \hat{D} \hat{F}$,
$\hat{F} \hat{D}^2 \hat{F}$ and $\hat{F}^3$ are possible.
The $\hat{F} \hat{D}^2 \hat{F}$ terms can be transformed
to the $\hat{D} \hat{F} \hat{D} \hat{F}$ terms
using the integration by parts
as in the case of ordinary gauge theory,
but $\hat{F}^3$ terms no longer vanish
even for the rank-one case.
Thus there are four terms at order $\alpha'$:\footnote{
Lorentz indices on the non-commutative side
should be regarded as being contracted using $G_{ij}$
although we will not write it explicitly
in this subsection
contrary to the conventions (\ref{implicit-g-1})
and (\ref{implicit-g-2}).
}
\begin{eqnarray}
&& \hat{T}_1 \equiv \hat{D}_i \hat{F}_{ik}
\ast \hat{D}_j \hat{F}_{jk}, \qquad
\hat{T}_2 \equiv \hat{D}_j \hat{F}_{ik}
\ast \hat{D}_i \hat{F}_{jk}, \qquad
\hat{T}_3 \equiv \hat{D}_k \hat{F}_{ij}
\ast \hat{D}_k \hat{F}_{ji},
\nonumber \\
&& \hat{T}_4 \equiv
i \hat{F}_{ij} \ast \hat{F}_{jk} \ast \hat{F}_{ki},
\end{eqnarray}
where we multiplied the $\hat{F}^3$ term by $i$
to make it Hermitian.
Using the Bianchi identity, the term $\hat{T}_3$
reduces to $\hat{T}_2$ as before,
\begin{equation}
\hat{T}_3 = -2 \hat{T}_2,
\end{equation}
but the terms $\hat{T}_1$ and $\hat{T}_2$
do not coincide up to total derivative since
\begin{eqnarray}
\hat{T}_1 &=& - \hat{F}_{ik} \ast \hat{D}_i \hat{D}_j \hat{F}_{jk}
+ {\rm total~derivative}, \\
\hat{T}_2 &=& - \hat{F}_{ik} \ast \hat{D}_j \hat{D}_i \hat{F}_{jk}
+ {\rm total~derivative},
\end{eqnarray}
where $\hat{D}_i$ and $\hat{D}_j$ no longer commute.
The remaining three terms $\hat{T}_1$, $\hat{T}_2$ and $\hat{T}_4$
are not independent which can be seen as follows:
\begin{eqnarray}
\hat{T}_1 - \hat{T}_2
&=& - \hat{F}_{ik} \ast [ \hat{D}_i, \hat{D}_j] \hat{F}_{jk}
+ {\rm total~derivative}
\nonumber \\
&=& - \hat{F}_{ik} \ast
(-i \hat{F}_{ij} \ast \hat{F}_{jk} +i \hat{F}_{jk} \ast \hat{F}_{ij})
+ {\rm total~derivative}
\nonumber \\
&=& -2 \hat{T}_4 + {\rm total~derivative}.
\end{eqnarray}
We will choose $\{ \hat{T}_1, \hat{T}_4 \}$
as a basis of $O(\alpha')$ terms in non-commutative gauge theory.

The origin of the extra term $\hat{T}_4$ can be interpreted
as an ambiguity in constructing non-commutative gauge theory from
ordinary gauge theory for the rank-one case.
This can be seen manifestly if we rewrite $\hat{T}_4$ as
\begin{equation}
\hat{T}_4 = \frac{i}{2} \hat{F}_{ij} \ast
(\hat{F}_{jk} \ast \hat{F}_{ki} - \hat{F}_{ki} \ast \hat{F}_{jk})
= \frac{1}{2} \hat{F}_{ij} \ast [ \hat{D}_k, \hat{D}_i] \hat{F}_{jk},
\end{equation}
which precisely corresponds to the ambiguity of the ordering
of covariant derivatives when we construct
a non-commutative counterpart of the term
$F_{ij} \partial_k \partial_i F_{jk}$.
This is characteristic of the rank-one theory and
there is no such ambiguity in higher-rank cases
where the ordering of field strengths or covariant derivatives
is already determined in ordinary Yang-Mills theory.

We have found bases of $O(\alpha')$ terms
for both ordinary and non-commutative gauge theories.
We will next consider the properties of the bases
with respect to their behavior under field redefinitions
and to the relation to our assumptions (\ref{two}).

For ordinary gauge theory our basis consists of $T_1$ alone.
It is possible to absorb $T_1$ into the $F^2$ term
by a field redefinition which is given by
\begin{eqnarray}
\tilde{A}_i &=& A_i
+ a (2 \pi \alpha') \partial_j F_{ji}
+ O(\alpha'^2), \\
\tilde{F}_{ij} \tilde{F}_{ji}
&=& F_{ij} F_{ji}
+ 4 a (2 \pi \alpha') \partial_i F_{ij} \partial_k F_{kj}
+ {\rm total~derivative} + O(\alpha'^2).
\end{eqnarray}
It is important to notice that this field redefinition has
the following property:
\begin{equation}
(B+ \tilde{F})_{ij}
= (B+F)_{ij} + a (2 \pi \alpha') \partial^2 (B+F)_{ij}
+ O(\alpha'^2).
\end{equation}
This implies that if the effective Lagrangian
in terms of $\tilde{A}_i$ depends on $B$ only in the form of $B+F$,
the Lagrangian in terms of $A_i$ also depends on $B$
only in the combination $B+F$, namely,
both $A_i$ and $\tilde{A}_i$ satisfy
the first assumption of (\ref{two}).
As can be seen from this example, the first assumption of (\ref{two})
does not determine the definition of the gauge field uniquely.
For instance, field redefinitions of the form
\begin{equation}
\tilde{A}_i = A_i
+ f_i (\partial F, \partial^2 F, \ldots),
\end{equation}
where field strengths in $f_i$
are accompanied by at least one derivative,
do not spoil the first assumption of (\ref{two}).
Since the term $T_1$ satisfies the initial term condition
(\ref{initial}) because of the fact that
$\partial_i (B+F)_{jk} = \partial_i F_{jk}$
for a constant $B$, we can proceed allowing a finite $T_1$ term
to be present in the Lagrangian without restricting
the definition of the gauge field further.
However we will take an alternative approach that
we choose a definition of the gauge field
in terms of which the $T_1$ term vanishes in the Lagrangian
among the ones
which satisfy
the first assumption of (\ref{two})
for convenience.

For non-commutative gauge theory our basis consists of
$\hat{T}_1$ and $\hat{T}_4$.
As in the case of ordinary gauge theory,
the term $\hat{T}_1$ can be absorbed into the $\hat{F}^2$ term
by a field redefinition given by
\begin{eqnarray}
\tilde{\hat{A}}_i &=& \hat{A}_i
+ a (2 \pi \alpha') \hat{D}_j \hat{F}_{ji}
+ O(\alpha'^2), \\
\tilde{\hat{F}}_{ij} \ast \tilde{\hat{F}}_{ji}
&=& \hat{F}_{ij} \ast \hat{F}_{ji}
+ 4 a (2 \pi \alpha') \hat{D}_i \hat{F}_{ij}
\ast \hat{D}_k \hat{F}_{kj}
+ {\rm total~derivative} + O(\alpha'^2).
\end{eqnarray}
This field redefinition preserves
the second assumption of (\ref{two})
so that we can select a definition of the non-commutative gauge
field satisfying (\ref{two})
such that the term $\hat{T}_1$ vanishes
in the Lagrangian.
With this convention and the one
for the ordinary gauge field we mentioned in the last paragraph,
there is no $O(\alpha')$ term
in ${\cal L} (B+F)$ and only the $\hat{T}_4$ term exists
in $\hat{{\cal L}} (\hat{F})$ at order $\alpha'$,
which implies that
\begin{equation}
\hat{A}_i = A_i + O(\alpha'^2),
\label{no-alpha'}
\end{equation}
namely, no $O(\alpha')$ part in the field redefinition.

On the other hand the term $\hat{T}_4$ cannot be redefined away
and it gives a non-vanishing contribution to the S-matrix
at $O(B)$ as we will see shortly.
It would be rather trivial that the existence of $\hat{T}_4$
in the effective Lagrangian is consistent with our assumptions
(\ref{two}) for the rank-one case
since it vanishes in the commutative limit.
Incidentally, the term $\hat{T}_4$ is consistent
for higher-rank cases as well
since its commutative counterpart $i \, {\rm tr} {\rm Tr} F^3$,
where tr denotes the trace over color indices,
satisfies the initial term condition (\ref{initial}),
which can be shown as follows:
\begin{eqnarray}
i \, {\rm tr} {\rm Tr} (B+F)^3 &=&
\frac{i}{2} {\rm tr} (B+F)_{ij} [ (B+F)_{jk}, (B+F)_{ki} ]
\nonumber \\
&=& \frac{i}{2} {\rm tr} F_{ij} [ F_{jk}, F_{ki} ]
+ \frac{i}{2} B_{ij} {\rm tr} [ F_{jk}, F_{ki} ]
\nonumber \\
&=& i \, {\rm tr} {\rm Tr} F^3.
\end{eqnarray}

\subsection{Constraints on two-derivative corrections}
\indent

In Section 3.1, we have shown that the $\hat{F}^2$ term
produces a non-vanishing contribution to the S-matrix
of order $O(B, \zeta^3, k^3)$ and that the $F^4$ terms are
determined by the requirement that
the Lagrangian ${\cal L} (B+F)$ should
reproduce the contribution.
Having understood that the term $\hat{T}_4$
is possible at order $\alpha'$, let us develop a similar discussion
for two-derivative corrections.

The term $\hat{T}_4$ is evaluated
in the $\alpha'$ expansion as follows:
\begin{equation}
\hat{T}_4 = \frac{1}{2} (2 \pi \alpha')^2 B_{nm}
\hat{F}_{ij} \partial_n \hat{F}_{jk} \partial_m \hat{F}_{ki}
+ O(\alpha'^4).
\label{T_4-expansion}
\end{equation}
We can extract the part which gives a non-vanishing contribution
to the three-point amplitude using the formula (\ref{formula}).
The result is
\begin{eqnarray}
\hat{T}_4 &=& (2 \pi \alpha')^2 B_{nm} \partial_i \hat{A}_j
\partial_n \partial_j \hat{A}_k \partial_m \partial_k \hat{A}_i
\nonumber \\
&& + {\rm ~terms~with~} \partial^2 A
+ {\rm total~derivative}
+ O(\alpha'^4).
\label{T_4-S}
\end{eqnarray}
The first term on the right-hand side of (\ref{T_4-S})
provides a non-vanishing contribution to the S-matrix
of order $O(B, \zeta^3, k^5)$.

On the commutative side, only terms of the form $O(\partial^2 F^4)$
can produce the same form of the contribution
after replacing $F$ with $B+F$.
Any term of order $O(\partial^2 F^4)$ can be
transformed to the following form
using the integration by parts
and the Bianchi identity \cite{AT1}:
\begin{equation}
{\cal L} = \sum_{i=1}^{7} b_i J_i,
\end{equation}
where
\begin{eqnarray}
&& J_1 = \partial_n F_{ij} \partial_n F_{ji} F_{kl} F_{lk},
\quad
J_2 = \partial_n F_{ij} \partial_n F_{jk} F_{kl} F_{li},
\nonumber \\
&& J_3 = F_{ni} F_{im} \partial_n F_{kl} \partial_m F_{lk},
\quad
J_4 = \partial_n F_{ni} \partial_m F_{im} F_{kl} F_{lk},
\nonumber \\
&& J_5 = -\partial_n F_{ni} \partial_m F_{ij} F_{jk} F_{km},
\quad
J_6 = \partial^2 F_{ij} F_{ji} F_{kl} F_{lk},
\nonumber \\
&& J_7 = \partial^2 F_{ij} F_{jk} F_{kl} F_{li},
\quad
\partial^2 F_{ij}
= \partial_i \partial_k F_{kj} - \partial_j \partial_k F_{ki}.
\label{AT}
\end{eqnarray}
The terms $J_4$, $J_5$, $J_6$ and $J_7$ contain the part
$\partial_j F_{ji}$ so that they do not contribute to the S-matrix.
This holds after replacing $F$ with $B+F$
since the part $\partial_j F_{ji}$ remains intact
in the replacement.
Thus we do not need to consider these terms
in the search for the term which reproduces the contribution from
the term $\hat{T}_4$.
On the other hand,
the first three coefficients $b_1$, $b_2$ and $b_3$
in this basis do not change
under field redefinition and unambiguous \cite{AT2}.
Therefore our goal is to answer the question whether
these coefficients are constrained by our assumptions (\ref{two}).

Let us denote the $O(B^n)$ part of $J_i$
with $F$ replaced by $B+F$ as $J_i (B^n)$ following \cite{Okawa}.
Explicit expressions of $J_i (B)$ and $J_i (B^2)$ for $i=1, 2, 3$
are given by
\begin{eqnarray}
&& J_1 (B) = 2 \partial_n F_{ij} \partial_n F_{ji} B_{kl} F_{lk},
\quad J_1 (B^2) = \partial_n F_{ij} \partial_n F_{ji} B_{kl} B_{lk},
\nonumber \\
&& J_2 (B) = 2 B_{ij} F_{jk} \partial_n F_{kl} \partial_n F_{li},
\quad J_2 (B^2) = \partial_n F_{ij} \partial_n F_{jk} B_{kl} B_{li},
\nonumber \\
&& J_3 (B) = 2 B_{ni} F_{im} \partial_n F_{kl} \partial_m F_{lk},
\quad J_3 (B^2) = B_{ni} B_{im} \partial_n F_{kl} \partial_m F_{lk}.
\end{eqnarray}
It is easily seen that the values of $J_1 (B^2)$, $J_2 (B^2)$
and $J_3 (B^2)$ vanish if they are evaluated
at on-shell configurations
(\ref{configuration}) satisfying (\ref{on-shell-condition}).
We can also show that the terms $J_1 (B)$ and $J_2 (B)$
do not contribute
to the S-matrix using the formula (\ref{formula}).
Therefore the term $J_3 (B)$ is
the only one which contributes to the S-matrix
of order $O(B, \zeta^3, k^5)$ on the commutative side,
which can be rewritten
using (\ref{formula}) as follows:
\begin{eqnarray}
J_3 (B) &=& 4 B_{ni} \partial_i A_m \partial_n \partial_k A_l
\partial_m \partial_l A_k + {\rm ~terms~with~} \partial^2 A
+ {\rm total~derivative}
\nonumber \\
&=& -4 B_{ni} \partial_i \partial_l A_m \partial_n \partial_k A_l
\partial_m A_k
\nonumber \\
&& + {\rm ~a~term~with~} \partial_l A_l
+ {\rm ~terms~with~} \partial^2 A
+ {\rm total~derivative}
\nonumber \\
&=& -4 B_{nm} \partial_i A_j
\partial_n \partial_j A_k \partial_m \partial_k A_i
\nonumber \\
&& + {\rm ~a~term~with~} \partial \cdot A
+ {\rm ~terms~with~} \partial^2 A
+ {\rm total~derivative}.
\end{eqnarray}
The non-vanishing contribution to the S-matrix from $J_3 (B)$
takes the same form as that of $\hat{T}_4$ (\ref{T_4-S})
so that it is possible to reproduce the S-matrix from $\hat{T}_4$
by $J_3 (B)$ with the following normalization factor:
\begin{equation}
\hat{T}_4 \sim -\frac{1}{4} (2 \pi \alpha')^2 J_3 (B).
\label{T_4-J_3}
\end{equation}

In addition to $J_3$, we can add the terms $J_1$ and $J_2$
to the effective Lagrangian
without violating the assumptions (\ref{two})
since $J_1 (B)$ and $J_2 (B)$ do not contribute to the S-matrix
at the order we are discussing.
In general, if a term $f(F)$ satisfies the condition that
\begin{equation}
f(B+F) = f(F) + {\rm total~derivative~using~the~equation~of~motion},
\label{on-shell-initial}
\end{equation}
we can add the term to the effective Lagrangian
without violating the assumptions (\ref{two}) at the same order
of $\alpha'$ as $f(F)$.
We will call (\ref{on-shell-initial})
the {\it on-shell initial term condition}.
Following this terminology, we can say that
the terms $J_1$ and $J_2$ do not satisfy the initial term condition
(\ref{initial}) but satisfy the on-shell initial term condition
(\ref{on-shell-initial}).

To summarize, the coefficients in front of $J_1$ and $J_2$
are not constrained by the assumptions (\ref{two})
since $J_1$ and $J_2$ satisfy the on-shell initial term condition.
The coefficient in front of $J_3$ is correlated with that in front of
$\hat{T}_4$ on the non-commutative side
following the relation (\ref{T_4-J_3}).
However, the coefficient in front of $\hat{T}_4$ was arbitrary
as we discussed in the preceding subsection so that
the coefficient in front of $J_3$ is also arbitrary.
Thus our conclusion is that two-derivative corrections of the form
$O(\partial^2 F^4)$ are not constrained at all
by the assumptions (\ref{two}) at this order.

This result may seem discouraging in view of our motivation
to obtain constraints on the effective Lagrangian.
However we do not expect that it holds at higher-order terms
in the $\alpha'$ expansion because of the following argument.
In general it would become more difficult
to satisfy the on-shell initial term condition
when the number of field strengths minus
the number of derivatives increases
in the term under consideration.
If we note that the existence of the solutions
to the on-shell initial term conditions was
essential to our conclusion that there is no constraint
on the $O(\partial^2 F^4)$ terms,
we can reasonably expect severe constraints
on such higher-order terms.
We admit, however, that the approach presented in this paper
will not be practical in deriving the constraints
on the higher-order terms
and we need more efficient methods.
As an example of promising methods
we can refer to the one discussed in \cite{Terashima}.
We will get back to this point after discussing the issue
on field redefinitions.

There is another comment on our result regarding
the relation between the coefficients in front of
$\hat{T}_4$ and $J_3$ (\ref{T_4-J_3}).
This provides no information on the effective Lagrangian
for the rank-one case since $\hat{T}_4$ vanishes
in the commutative limit.
However if we succeed in extending our consideration
to higher-rank cases, it might be possible to obtain
a prediction on a relation between the coefficient in front of
the $F^3$ term and coefficients in $O(D^2 F^4)$ terms.

We should now clarify the relation between the result in this paper
and that in \cite{Okawa}.
The most general form of $O(\partial^2 F^4)$ terms
was derived in \cite{Okawa} from the requirement that
${\cal L} (B+F)$ and $\hat{{\cal L}} (\hat{F})$ coincide
up to total derivative under the assumption that
the field redefinition is given by (\ref{A-hat-2}).
The result was that the terms $J_1$, $J_2$ and $J_3$ must appear
in the combination that
\begin{equation}
-\frac{1}{4} J_1 +2 J_2 + J_3.
\label{Okawa-J}
\end{equation}
This was inconsistent with
the $O(\partial^2 F^4)$ terms
in bosonic string theory derived from the string four-point
amplitude \cite{AT1} or from
the two-loop beta function
in the open string sigma model \cite{AT2}
which are proportional to\footnote{
This expression is slightly different from
(4) in \cite{AT2} but one of
the authors was informed of
a misprint in (4) of \cite{AT2}:
the last coefficient $b_3$ should have sign $+$.
}
\begin{equation}
-\frac{1}{4} J_1 -2 J_2 + J_3.
\label{AT-J}
\end{equation}
The conclusion in this paper that no constraint is imposed
on $O(\partial^2 F^4)$ terms
is trivially consistent with (\ref{AT-J})
and the difference between this conclusion
and that in \cite{Okawa} implies that
the relation between
the two gauge fields $\hat{A}_i$ and $A_i$ in (\ref{two})
does not in general take the form of (\ref{A-hat-2})
assumed in \cite{Okawa}.
In particular, the discrepancy
between (\ref{Okawa-J}) and (\ref{AT-J})
shows that it is indeed the case for bosonic string theory.
We will construct a field redefinition which is relevant
to bosonic string theory in the next subsection.

\subsection{Corrections to the field redefinition}
\indent

We presented the on-shell initial term condition
(\ref{on-shell-initial}) as a necessary condition
for a term to be added to the effective Lagrangian
without violating the assumptions (\ref{two})
in the preceding subsection.
The relation between $\hat{A}_i$ and $A_i$
must be in general modified
if we add a term which satisfies the on-shell
initial term condition (\ref{on-shell-initial}) but does not
satisfy the initial term condition (\ref{initial}).
As we have seen, the terms $J_1$ and $J_2$
are examples of such terms
since $J_1 (B)$, $J_1 (B^2)$, $J_2 (B)$ and $J_2 (B^2)$
are not total derivative although values of them vanish
when evaluated at configurations satisfying the on-shell
conditions (\ref{on-shell-condition}).
The terms $J_4$, $J_5$, $J_6$ and $J_7$
also satisfy the on-shell initial
term condition, however, they are less interesting than $J_1$ and
$J_2$ since they do not contribute to the S-matrix.
An explicit form of the required field redefinition
which relates $\hat{A}_i$ to $A_i$ when we add a term
which satisfies the on-shell initial term condition
to the effective Lagrangian
can be derived in the same way as we did in Section 3.2
but we will not do that for completely general cases.
It would be sufficient to demonstrate it for some examples
including the one which is relevant to bosonic string theory
since the generalization is straightforward.

Let us first consider a case where only $J_2$ exists
in the $O(\alpha'^3)$ part.
In particular, the absence of $J_3$ means that
$\hat{T}_4$ is not allowed to exist in $\hat{{\cal L}} (\hat{F})$
because of the relation (\ref{T_4-J_3}).
Thus there are no $O(\alpha')$ terms in $\hat{{\cal L}} (\hat{F})$
under our convention that $\hat{T}_1$ should be redefined away.
This simplifies the discussion since the $O(\alpha'^2)$ part 
in the field redefinition (\ref{redefinition-1}),
which is necessary
to satisfy the assumptions (\ref{two}) as we have seen
in the preceding section,
does not affect $O(\alpha'^3)$ terms under consideration
if there are no $O(\alpha')$ terms in $\hat{{\cal L}} (\hat{F})$.
Furthermore, the $O(\alpha'^2)$ part
of $\hat{{\cal L}} (\hat{F})$ cannot generate $B$-dependent terms
of order $\alpha'^3$
which is manifest under our convention (\ref{no-alpha'}).
Therefore the terms $J_2 (B)$ and $J_2 (B^2)$, which are necessary
to realize the $B$-dependence of the form $B+F$
when we add $J_2$, must be generated from the $\hat{F}^2$ term
by the $O(\alpha'^3)$ part of the field redefinition of $\hat{A}_i$.
Its explicit form is easily derived if we rewrite
$J_2 (B)$ and $J_2 (B^2)$ as follows:
\begin{eqnarray}
J_2 (B) &=&
-2 \partial_n F_{ni} \partial_j (F B F)_{ji}
+ {\rm total~derivative},
\\
J_2 (B^2) &=&
- \partial_n F_{ni} \partial_j (B^2 F + F B^2)_{ji}
+ {\rm total~derivative}.
\end{eqnarray}
It follows from a similar argument to the one used
to determine the form (\ref{redefinition-1}) that
the field redefinition
\begin{eqnarray}
\hat{A}_i &=& A_i + \frac{1}{2} (2 \pi \alpha')^2
B_{kl} A_k (\partial_l A_i + F_{li})
\nonumber \\
&& -\frac{1}{4} c_2 (2 \pi \alpha')^3
\partial_j ( 2 F B F + B^2 F + F B^2 )_{ji}
+ O(\alpha'^4)
\label{redefinition-2}
\end{eqnarray}
generates $c_2 (2 \pi \alpha')^3 ( J_2 (B) + J_2 (B^2) )$ from
the $\hat{F}^2$ term.
To summarize, the two Lagrangians,
\begin{eqnarray}
{\cal L}(B+F) &=& \frac{\sqrt{\det g}}{g_s} \Biggl[
{\rm Tr} (B+F)^2
+ (2 \pi \alpha')^2 \left[
\frac{1}{2} {\rm Tr} (B+F)^4
-\frac{1}{8} [ {\rm Tr} (B+F)^2 ]^2 \right]
\nonumber \\
&& \qquad \qquad + c_2 (2 \pi \alpha')^3
\partial_n (B+F)_{ij} \partial_n (B+F)_{jk} (B+F)_{kl} (B+F)_{li}
\nonumber \\
&& \qquad \qquad + ~O(\alpha'^4)
\Biggr],
\end{eqnarray}
and
\begin{eqnarray}
\hat{{\cal L}}(\hat{F}) &=& \frac{\sqrt{\det G}}{G_s} \Biggl[
{\rm Tr} ( G^{-1} \hat{F} \ast G^{-1} \hat{F} )
\nonumber \\
&& \qquad \quad + (2 \pi \alpha')^2 \biggl[
\frac{1}{2} {\rm Tr} ( G^{-1} \hat{F} )^4_{\rm arbitrary}
-\frac{1}{8} ( {\rm Tr} ( G^{-1} \hat{F} )^2 )^2_{\rm arbitrary}
\biggr]
\nonumber \\
&& \qquad \quad + c_2 (2 \pi \alpha')^3
( \hat{D}_n \hat{F}_{ij} \ast \hat{D}_n \hat{F}_{jk}
\ast \hat{F}_{kl} \ast \hat{F}_{li} )_{G, \rm~arbitrary}
+ ~O(\alpha'^4)
\Biggr],
\end{eqnarray}
with an arbitrary ordering of $\hat{D} \hat{F}$'s and $\hat{F}$'s
in the $O(\alpha'^3)$ term contracted using $G_{ij}$
as indicated by the subscript,
are related by the field redefinition (\ref{redefinition-2}).

This example shows that
$\alpha'$ corrections of $O(B)$ to the field redefinition
(\ref{redefinition-1}) are in general possible.
Since
\begin{equation}
B_{ij} = -\frac{1}{(2 \pi \alpha')^2}
( g \theta g )_{ij} + O( \theta^2 ),
\end{equation}
this does not take the form of (\ref{A-hat-2}).
Therefore it would be helpful to confirm that (\ref{redefinition-2})
preserves the gauge equivalence relation (\ref{A-hat}).
Let us decompose the field redefinition (\ref{redefinition-2})
as follows:
\begin{equation}
\hat{A}_i \to \tilde{A}_i \to A_i,
\end{equation}
where
\begin{eqnarray}
\hat{A}_i &=& \tilde{A}_i + \frac{1}{2} (2 \pi \alpha')^2
B_{kl} \tilde{A}_k (\partial_l \tilde{A}_i + \tilde{F}_{li})
+ O(\alpha'^4),
\label{redefinition-2-1} \\
\tilde{A}_i &=& A_i 
-\frac{1}{4} c_2 (2 \pi \alpha')^3
\partial_j ( 2 F B F + B^2 F + F B^2 )_{ji}
+ O(\alpha'^4).
\label{redefinition-2-2}
\end{eqnarray}
By the first part (\ref{redefinition-2-1}),
the non-commutative gauge field $\hat{A}_i$ is mapped to
an ordinary gauge field $\tilde{A}_i$ which respects the ordinary
gauge invariance
while $\tilde{A}_i$ is mapped to another ordinary gauge field $A_i$
by the second part (\ref{redefinition-2-2})
since the difference between $\tilde{A}_i$ and $A_i$
is gauge invariant although it depends on $B$.
This shows that (\ref{redefinition-2})
preserves the gauge equivalence relation (\ref{A-hat}).
In general, the field redefinition (\ref{redefinition-1})
maps a non-commutative gauge field to an ordinary gauge field
but the $B$-dependence of the effective Lagrangian in terms of
the resulting gauge field, $\tilde{A}_i$ in this example,
does not take the form of $B+F$.
Therefore further $B$-dependent redefinition like
(\ref{redefinition-2-2})
is necessary to map it to the gauge field which satisfies
the first assumption of (\ref{two}).

The form of the field redefinition (\ref{redefinition-2})
does not belong to the class of solutions
to the gauge equivalence relation
(\ref{A-hat}) found in \cite{AK}.
However there is no contradiction
since it was assumed in \cite{AK} that
Lorentz indices in a mapping from $A_i$ to $\hat{A}_i$
are contracted among derivatives of the gauge field
and $\delta \theta^{ij}$ alone
while $(g^{-1})^{ij}$ is used in our case (\ref{redefinition-2})
although it is implicit
under our convention (\ref{implicit-g-1}).\footnote{
We would like to thank I. Kishimoto for clarifying this point.
}

Now the extension to cases where other $J_i$'s except $J_3$
exist in the effective Lagrangian would be straightforward.
However if $J_3$ exists the story becomes slightly complicated
because of the presence of $\hat{T}_4$ in $\hat{{\cal L}} (\hat{F})$
which accompanies $J_3$ following the relation (\ref{T_4-J_3}).
We have to consider the effects of the $O(\alpha'^2)$ part
in the field redefinition (\ref{redefinition-1})
when it acts on the $O(\alpha')$ term $\hat{T}_4$.
Here it is convenient to utilize the results of \cite{Okawa}.
Let us review them briefly.

It was shown in \cite{Okawa} that the two Lagrangians,
\begin{equation}
\hat{{\cal L}}_1 (\hat{F}) =
\frac{\sqrt{\det G}}{G_s} \left[
\hat{T}_3 + (2 \pi \alpha')^2
\left( -\frac{1}{4} \hat{J}_1 + 2 \hat{J}_2 + \hat{J}_3
\right) + O(\alpha'^4)
\right],
\end{equation}
and
\begin{equation}
\hat{{\cal L}}_2 (\hat{F}) =
\frac{\sqrt{\det G}}{G_s} \left[
\hat{T}_1 + (2 \pi \alpha')^2
\left( \hat{J}_5 -\frac{1}{8} \hat{J}_6 + \frac{1}{2} \hat{J}_7
\right) + O(\alpha'^4)
\right],
\end{equation}
satisfy $\hat{{\cal L}} (\hat{F}) = {\cal L} (B+F)$
up to total derivative under the field redefinition
(\ref{redefinition-1}) with the definition of $G_s$ (\ref{G_s}).
Here $\hat{J}_i$'s are the non-commutative counterparts of
$J_i$'s with an arbitrary ordering of the fields.
We presented the Lagrangians on the non-commutative side
because we can uniquely construct their commutative counterparts
while the other direction,
${\cal L} (B+F) \to \hat{{\cal L}} (\hat{F})$, suffers from
the ambiguity in the rank-one case discussed in Section 4.1.
A linear combination of the two Lagrangians is expressed
in our basis $\{ \hat{T}_1, \hat{T}_4 \}$ as follows:
\begin{eqnarray}
\hat{{\cal L}} (\hat{F}) &=&
\frac{\sqrt{\det G}}{G_s} \Biggl[
a \hat{T}_1 + b \hat{T}_4
+ a (2 \pi \alpha')^2
\left( \hat{J}_5 -\frac{1}{8} \hat{J}_6 + \frac{1}{2} \hat{J}_7
\right)
\nonumber \\
&& \quad
- \frac{1}{4} b (2 \pi \alpha')^2
\left( -\frac{1}{4} \hat{J}_1 + 2 \hat{J}_2 + \hat{J}_3
+2 \hat{J}_5 -\frac{1}{4} \hat{J}_6 + \hat{J}_7
\right) + O(\alpha'^4) \Biggr].
\label{general-form}
\end{eqnarray}
It was further argued in \cite{Okawa} that
(\ref{general-form}) is the most general form
of two-derivative corrections
up to this order in the $\alpha'$ expansion
which satisfy
$\hat{{\cal L}} (\hat{F}) = {\cal L} (B+F)$
up to total derivative under the field redefinition
(\ref{redefinition-1})
with the definition of $G_s$ (\ref{G_s}).\footnote{
The argument for proving this statement
developed in Section 3 of \cite{Okawa}
was incorrect as explained in the note added
at the end of hep-th/9909132 v2.
}
To see that it is the case, it is helpful to notice that
if there is another Lagrangian
\begin{eqnarray}
\hat{{\cal L}}' (\hat{F}) &=&
\frac{\sqrt{\det G}}{G_s} \Biggl[
a \hat{T}_1 + b \hat{T}_4
\nonumber \\
&& + ~O(\alpha'^2) {\rm~terms~different~from~those~of~}
\hat{{\cal L}} (\hat{F})
+ O(\alpha'^4) \Biggr],
\end{eqnarray}
which also satisfies
$\hat{{\cal L}}' (\hat{F}) = {\cal L}' (B+F)$
up to total derivative under the field redefinition
(\ref{redefinition-1}) with the definition of $G_s$ (\ref{G_s}),
then the difference
${\cal L}' (F) - {\cal L} (F)$
must be a solution of the form $O(\partial^2 F^4)$
to the initial term condition (\ref{initial}).
Thus the question whether (\ref{general-form}) is
the most general form reduces to the one
whether there are solutions of the form $O(\partial^2 F^4)$
to the initial term condition (\ref{initial}).
Regarding the latter question,
it was shown \cite{Okawa} that the condition that
$O(\partial^2 F^4)$ terms must be proportional to
the combination that
\begin{equation}
{\cal F} (F) \equiv -\frac{1}{4} J_1 + 2 J_2 + J_3
+2 J_5 -\frac{1}{4} J_6 + J_7,
\label{necessary}
\end{equation}
is necessary
to satisfy the initial term condition (\ref{initial}).
It is difficult to see whether
${\cal F} (F)$ satisfies the initial term condition
by a direct calculation, however, we can obtain the answer
by an indirect argument in the following way. From
(\ref{T_4-expansion}) and the fact that
(\ref{general-form}) with $a=0$ satisfies
$\hat{{\cal L}} (\hat{F}) = {\cal L} (B+F)$
up to total derivative under the field redefinition
(\ref{redefinition-1}) with the definition of $G_s$ (\ref{G_s}),
it follows that
\begin{equation}
-\frac{1}{4} {\cal F} (B+F) =
\frac{1}{2} B_{nm}
F_{ij} \partial_n F_{jk} \partial_m F_{ki}
-\frac{1}{4} {\cal F} (F)
+ {\rm total~derivative},
\end{equation}
which implies that ${\cal F} (F)$ does not satisfy
the initial term condition (\ref{initial}).
Now that the only remaining possibility was denied,
the statement that
there is no solution of the form $O(\partial^2 F^4)$
to the initial term condition (\ref{initial})
was shown and this implies
that (\ref{general-form}) is the most general form
of two-derivative corrections
up to this order in the $\alpha'$ expansion
which satisfy
$\hat{{\cal L}} (\hat{F}) = {\cal L} (B+F)$
up to total derivative under the field redefinition
(\ref{redefinition-1}) with the definition of $G_s$ (\ref{G_s}).

This result provides us with a good starting point for the case
where $J_3$ is non-vanishing.
Namely, the Lagrangian
\begin{eqnarray}
\hat{{\cal L}}(\hat{F}) &=& \frac{\sqrt{\det G}}{G_s} \Biggl[
{\rm Tr} ( G^{-1} \hat{F} \ast G^{-1} \hat{F} )
+ b (2 \pi \alpha') \hat{T}_4
\nonumber \\
&& \qquad \qquad + (2 \pi \alpha')^2 \biggl[
\frac{1}{2} {\rm Tr} ( G^{-1} \hat{F} )^4_{\rm arbitrary}
-\frac{1}{8} ( {\rm Tr} ( G^{-1} \hat{F} )^2 )^2_{\rm arbitrary}
\biggr]
\nonumber \\
&& \qquad \qquad - \frac{1}{4} b (2 \pi \alpha')^3
\left( -\frac{1}{4} \hat{J}_1 + 2 \hat{J}_2 + \hat{J}_3
+2 \hat{J}_5 -\frac{1}{4} \hat{J}_6 + \hat{J}_7
\right)
\nonumber \\
&& \qquad \qquad + ~O(\alpha'^4)
\Biggr],
\label{non-vanishing-J_3}
\end{eqnarray}
and ${\cal L} (B+F)$ constructed from ${\hat{\cal L}} (\hat{F})$
are related by the field redefinition (\ref{redefinition-1}).
If we want to 
change the coefficients in font of $J_i$'s except $J_3$,
we should
modify the form of the field redefinition
at order $\alpha'^3$ appropriately
as in the preceding example where only $J_2$ exists.

As an interesting example of such cases,
let us derive the form of the field redefinition
which is relevant to bosonic string theory.
As we mentioned in the preceding subsection,
the coefficients in front of $J_1$, $J_2$ and $J_3$
calculated in bosonic string theory are
proportional to (\ref{AT-J}) \cite{AT1,AT2}.
This corresponds to adding $b (2 \pi \alpha')^3 \hat{J_2}$
to (\ref{non-vanishing-J_3}) so that the form of
the field redefinition is modified to
\begin{eqnarray}
\hat{A}_i &=& A_i + \frac{1}{2} (2 \pi \alpha')^2
B_{kl} A_k (\partial_l A_i + F_{li})
\nonumber \\
&& -\frac{1}{4} b (2 \pi \alpha')^3
\partial_j ( 2 F B F + B^2 F + F B^2 )_{ji}
+ O(\alpha'^4).
\label{redefinition-3}
\end{eqnarray}
If we further change the coefficients in front of $J_4$, $J_5$,
$J_6$ and $J_7$ which do not affect the S-matrix,
the form of the field redefinition (\ref{redefinition-3})
itself is modified correspondingly.
However we cannot make the $O(\alpha'^3)$ terms vanish
since (\ref{AT-J}) does not take the general form
(\ref{general-form}) in the absence of the $O(\alpha'^3)$ terms.
Thus the corrections to the field redefinition
(\ref{redefinition-1}) are not only possible in principle
but also realized actually in bosonic string theory.
For superstring theory, it was found that the coefficients
in front of $J_1$, $J_2$ and $J_3$ vanish \cite{AT1}
so that corrections to the field redefinition
(\ref{redefinition-1}) at order $\alpha'^3$
are not required.
However there is no general argument that it persists
to higher orders in the $\alpha'$ expansion.
We should keep such possibility of corrections in mind
when we use properties
of the field redefinition which relates the non-commutative
gauge field to the ordinary one.
In particular, it would be important to note that
corrections of $O(B) \sim O(\theta)$
modify the differential equation of $\delta \hat{A} (\theta)$
introduced in \cite{SW} for more general descriptions
of the system in terms of non-commutative gauge theory.

\section{Conclusions and discussions}
\setcounter{equation}{0}
\indent

We considered the constraints on the effective Lagrangian
of the gauge field on a single D-brane in flat space-time
imposed by the compatibility of the description
by non-commutative gauge theory $\hat{{\cal L}} (\hat{F})$
with that by ordinary gauge theory ${\cal L} (B+F)$
in the presence of a constant $B$ field background.
We presented a systematic method
under the $\alpha'$ expansion to derive
the constraints based on the assumptions (\ref{two}) alone
without assuming the form of the field redefinition
which relates the non-commutative gauge field $\hat{A}_i$
to the ordinary one $A_i$.

By applying this method to two-derivative corrections
to the DBI Lagrangian,
we established the equivalence of the two descriptions
for a larger class of Lagrangians.
In particular it contains the effective Lagrangian of bosonic
string theory and thus
the puzzle in bosonic strings
found in the previous works was resolved
and the equivalence in this case was first made consistent.

In resolving the puzzle it was essential to observe
that the gauge-invariant but $B$-dependent corrections
to the field redefinition (\ref{A-hat-2})
are in general necessary for the compatibility.
They were not considered previously because it was assumed that
the metric $g_{ij}$ does not appear in the field redefinition
which relates $\hat{A}_i$ to $A_i$,
however we showed
that they must exist for the case of bosonic string theory.
It should be emphasized that
it was crucial to reach this observation that
we did not assume the form of the field redefinition
when we derive the constraints.

It is sometimes said that the form of the field redefinition
which relates $\hat{A}_i$ to $A_i$
can be determined by solving the differential equation
derived from the gauge equivalence relation
in \cite{SW} up to the ambiguities found in \cite{AK}.
However our result clearly shows that
it is no longer the case if we allow the metric $g_{ij}$
to appear in the field redefinition
as in the case of bosonic string theory.
Even the form of the differential equation itself can be modified
and the form of the field redefinition is hardly constrained
by the gauge equivalence relation without the assumption.
In the superstring case the field redefinition of the form
(\ref{A-hat-2}) can be consistent
up to two derivatives \cite{Terashima}
and may not be corrected.
However, as far as we are aware of,
there is no argument which justifies the assumption
in the superstring case
that the metric $g_{ij}$ does not appear in the field redefinition
which relates $\hat{A}_i$ to $A_i$.

We believe that we have elucidated the mechanism to constrain
the effective Lagrangian of the gauge fields on D-branes using
non-commutative gauge theory.
Since we presented a systematic method to obtain
the constraints in the $\alpha'$ expansion,
it is in principle possible to calculate
the general form of the effective Lagrangian which satisfies
the assumptions (\ref{two})
up to an arbitrary order in $\alpha'$.
Furthermore our study implies that
the number of the free parameters in the general form
is equal to or less than the number of solutions to
the initial term condition (\ref{initial})
plus that of nontrivial solutions
to the on-shell initial term condition (\ref{on-shell-initial}).
Therefore if there are no solutions to these conditions
in the two-derivative terms at higher order in $\alpha'$, for example,
this implies that
the form of the two-derivative terms is in principle
determined uniquely by the requirement of the compatibility
up to the parameters in front of $J_1$, $J_2$ and $J_3$.
However we admit that the approach adopted in the present paper
is not practically useful to proceed to the higher orders
to obtain the constraints
or the explicit form of the terms
as we mentioned in Section 4.2.
Regarding this issue,
a general method to construct
$2n$-derivative terms to all orders in $\alpha'$
which satisfy the compatibility of the two descriptions
in the approximation of neglecting $(2n+2)$-derivative terms
when the field redefinition takes the form of (\ref{A-hat-2})
was presented in \cite{Terashima}.
It is therefore necessary to extend the method
to apply it to more general cases
where there are corrections to
the field redefinition of the form (\ref{A-hat-2})
such as the case of bosonic string theory.
If we could succeed in such generalization,
it would be expected that it will provide us with
a new powerful method to study the dynamics of D-branes.
For developments in this direction,
the simplified Seiberg-Witten map considered in
\cite{Cornalba-Chiappa} and \cite{Ishibashi}
may be useful
because of its geometric nature
although we should clarify its meaning for our approach.
See also a related work \cite{Okuyama}.
How to construct actions which are invariant
under the simplified map was recently discussed in \cite{Cornalba}.
In addition, it is interesting to combine our approach
with consideration of supersymmetry and string dualities.
It will probably provide us with further constraints.

We only considered the constraints on the effective Lagrangian
at the lowest order in the expansion
with respect to the string coupling constant $g_s$
in this paper.
There seems to be no crucial obstruction
to the extension of our approach to higher orders in $g_s$
although some modifications may be required.
An issue related to this kind of extension was discussed
in \cite{Andreev}.
Furthermore, although the assumptions (\ref{two}) were
derived from the action of the sigma model,
they are not related to the expansions with respect to
$\alpha'$ and $g_s$ once extracted.
It would be interesting
if we could obtain some non-perturbative information
on the dynamics of D-branes from them.
Of course it might be the case that
there are limitations of the description
in terms of non-commutative gauge theory
at non-perturbative level
and it is important to investigate them.

Another important extension of our approach
is to consider higher-rank gauge theory.
It would be interesting if we could obtain
some insight into the non-Abelian generalization
of the DBI Lagrangian \cite{NDBI}.
Although we foresee possible complication
originated in its non-Abelian nature
which exists even on the side of ordinary gauge theory,
it will be worth investigating
in view of the various important developments which have been made
by the super Yang-Mills theory
in the description of multi-body systems of D-branes.

\vspace{0.4cm}
\noindent
Acknowledgements

We would like to thank K. Hashimoto, H. Kajiura and T. Kawano
for useful communications on non-commutative gauge theory.
Y.O. also thanks N. Ohta, M. Sato and T. Yoneya
for valuable discussions and comments. 
This works of Y.O. and S.T. were supported in part
by the Japan Society for the Promotion of Science
under the Postdoctoral Research Programs No. 11-01732
and No. 11-08864 respectively.

\newpage
\appendix
\renewcommand{\thesection}{Appendix \Alph{section}.}
\renewcommand{\theequation}{\Alph{section}.\arabic{equation}}

\section{Determination of the $O(\alpha'^0)$ part of $G_s$}
\setcounter{equation}{0}
\indent

In this appendix, we determine the $O(\alpha'^0)$ part of $G_s$,
namely the constant $t$ in (\ref{G_s-0}), by the assumptions
(\ref{two}).
Since the normalizations of $A_i$ and $\hat{A}_i$
do not coincide when $t \ne 1$ as can be seen from (\ref{U1}),
we should be careful when evaluating the S-matrix.
A safer approach is to make calculations after
rescaling both fields such that their normalizations coincide.
We denote the normalized fields
by ${\bf A}_i$ and ${\bf \hat{A}}_i$:
\begin{equation}
{\bf A}_i = \frac{A_i}{\sqrt{g_s}}, \quad
{\bf \hat{A}}_i = \frac{\hat{A}_i}{\sqrt{G_s}}.
\end{equation}
If we define the field strengths of the normalized fields as
\begin{eqnarray}
{\bf F}_{ij} &\equiv&
\partial_i {\bf A}_j - \partial_j {\bf A}_i, \\
{\bf \hat{F}}_{ij} &\equiv&
\partial_i {\bf \hat{A}}_j - \partial_j {\bf \hat{A}}_i
-i \sqrt{G_s} {\bf \hat{A}}_i \ast {\bf \hat{A}}_j
+i \sqrt{G_s} {\bf \hat{A}}_j \ast {\bf \hat{A}}_i,
\end{eqnarray}
the effective Lagrangians ${\cal L} (B+F)$ (\ref{(B+F)^2})
and $\hat{{\cal L}} (\hat{F})$ (\ref{F-hat^2})
can be rewritten as follows:
\begin{eqnarray}
{\cal L} (B+F) &=&
\sqrt{\det g} \, {\rm Tr}
\left( \frac{B}{\sqrt{g_s}} + {\bf F} \right)^2
+ O(\alpha') \nonumber \\
&=& \sqrt{\det g} \, {\rm Tr} \, {\bf F}^2
+ {\rm total~derivative} + O(\alpha'), \\
\hat{{\cal L}} (\hat{F}) &=&
\sqrt{\det G} \, {\rm Tr}
( G^{-1} {\bf \hat{F}} \ast G^{-1} {\bf \hat{F}} )
+ O(\alpha').
\end{eqnarray}
It is clear from these expressions that
the normalized fields ${\bf A}_i$ and ${\bf \hat{A}}_i$
coincide at the lowest order in $\alpha'$:
\begin{equation}
{\bf \hat{A}}_i = {\bf A}_i + O(\alpha').
\end{equation}

Following the procedure presented in Section 3.1,
the $F^4$ terms can be determined in this case as well.
The evaluation of the Lagrangian on the non-commutative side
in the $\alpha'$ expansion is given 
in terms of the normalized field ${\bf \hat{A}}_i$ by
\begin{eqnarray}
&& \sqrt{\det G} \,
{\rm Tr } (G^{-1} {\bf \hat{F}} \ast G^{-1} {\bf \hat{F}})
\nonumber \\
&=& \sqrt{\det g} \Biggl[
( \partial_i {\bf \hat{A}}_j  - \partial_j {\bf \hat{A}}_i )
( \partial_j {\bf \hat{A}}_i  - \partial_i {\bf \hat{A}}_j )
-4 (2 \pi \alpha')^2 \sqrt{G_s} B_{kl}
\partial_k {\bf \hat{A}}_i
\partial_l {\bf \hat{A}}_j \partial_j {\bf \hat{A}}_i
\nonumber \\
&& +2 (2 \pi \alpha')^2 (B^2)_{ij}
( \partial_j {\bf \hat{A}}_k  - \partial_k {\bf \hat{A}}_j )
( \partial_k {\bf \hat{A}}_i  - \partial_i {\bf \hat{A}}_k )
\nonumber \\
&& -\frac{1}{2} (2 \pi \alpha')^2 {\rm Tr} B^2
( \partial_i {\bf \hat{A}}_j  - \partial_j {\bf \hat{A}}_i )
( \partial_j {\bf \hat{A}}_i  - \partial_i {\bf \hat{A}}_j )
+ O(\alpha'^4) \Biggr].
\label{L-hat-t}
\end{eqnarray}
The relevant terms on the commutative side are
\begin{eqnarray}
&& \frac{\sqrt{\det g}}{g_s} {\rm Tr} (B+F)^4
= \sqrt{\det g} \, g_s {\rm Tr} \left(
\frac{B}{\sqrt{g_s}} + {\bf F} \right)^4
\nonumber \\
&=& \sqrt{\det g} \, \left[
g_s {\rm Tr} \, {\bf F}^4
+ 4 \sqrt{g_s} \, {\rm Tr} B {\bf F}^3 + O(B^2)
\right], \\
&& \frac{\sqrt{\det g}}{g_s} [{\rm Tr} (B+F)^2]^2
= \sqrt{\det g} \, g_s \left[ {\rm Tr} \left(
\frac{B}{\sqrt{g_s}} + {\bf F} \right)^2 \right]^2
\nonumber \\
&=& \sqrt{\det g} \, \left[
g_s ( {\rm Tr} \, {\bf F}^2 )^2
+ 4 \sqrt{g_s} \,
{\rm Tr} B {\bf F} \, {\rm Tr} \, {\bf F}^2 + O(B^2)
\right].
\end{eqnarray}
The requirement that both Lagrangians $\hat{{\cal L}} (\hat{F})$
and ${\cal L} (B+F)$ should produce
the same S-matrix of order $O(B,\zeta^3,k^3)$ determines
the form of the Lagrangian ${\cal L} (B+F)$
in a completely parallel way to Section 3.1.
The result is as follows:
\begin{eqnarray}
{\cal L} (B+F) &=&
\sqrt{\det g} \left[
{\rm Tr}
\left( \frac{B}{\sqrt{g_s}} + {\bf F} \right)^2
+ (2 \pi \alpha')^2 \sqrt{G_s \, g_s}
\left[
\frac{1}{2} {\rm Tr} \left(
\frac{B}{\sqrt{g_s}} + {\bf F} \right)^4
\right. \right. \nonumber \\ && \qquad \qquad \left. \left.
-\frac{1}{8} \left( {\rm Tr} \left(
\frac{B}{\sqrt{g_s}} + {\bf F} \right)^2 \right)^2
\right] + O(\alpha'^4) + {\rm derivative~corrections} \right]
\nonumber \\
&=& \frac{\sqrt{\det g}}{g_s} \Biggl[
{\rm Tr} (B+F)^2
+ \sqrt{t} \, (2 \pi \alpha')^2 \left[
\frac{1}{2} {\rm Tr} (B+F)^4
\right. \nonumber \\ && \qquad \qquad \left.
-\frac{1}{8} [ {\rm Tr} (B+F)^2 ]^2 \right]
+ ~O(\alpha'^4) + {\rm derivative~corrections}
\Biggr],
\label{L-t}
\end{eqnarray}
where we used $t$ defined by (\ref{G_s-0}).
The $O(B)$ part of this Lagrangian coincides with that of
(\ref{L-hat-t}) up to field redefinition since
both produce the same S-matrix,
but it may not the case for the $O(B^2)$ part.
The $O(\alpha'^2)$ part of
the Lagrangian (\ref{L-t}) expanded with respect to $B$
is given by
\begin{eqnarray}
&& \sqrt{\det g} \, (2 \pi \alpha')^2 \sqrt{G_s \, g_s}
\left[
\frac{1}{2} {\rm Tr} \left(
\frac{B}{\sqrt{g_s}} + {\bf F} \right)^4
-\frac{1}{8} \left( {\rm Tr} \left(
\frac{B}{\sqrt{g_s}} + {\bf F} \right)^2 \right)^2
\right]
\nonumber \\
&=& \sqrt{\det g} \, (2 \pi \alpha')^2 \Biggl[
\sqrt{t} \, g_s \left[
\frac{1}{2} {\rm Tr} \, {\bf F}^4
-\frac{1}{8} ( {\rm Tr} \, {\bf F}^2 )^2 \right]
\nonumber \\
&& + \sqrt{t \, g_s} \left(
2 {\rm Tr} B {\bf F}^3
-\frac{1}{2} {\rm Tr} B {\bf F} \, {\rm Tr} \, {\bf F}^2
\right)
\nonumber \\
&& + \sqrt{t} \left(
2 {\rm Tr} B^2 {\bf F}^2
-\frac{1}{4} {\rm Tr} B^2 \, {\rm Tr} \, {\bf F}^2
\right)
+ {\rm total~derivative}
+ {\rm const.} \Biggr].
\label{(B+F)^4-t}
\end{eqnarray}
The $O(B^2)$ part does not coincide with that of (\ref{L-hat-t}).
The difference in the ${\rm Tr} B^2 {\rm Tr} F^2$ term
can be absorbed by field redefinition as we explained
in Section 3.3
so that it is irrelevant to the determination of $t$,
whereas the values of the ${\rm Tr} B^2 F^2$ term
evaluated at on-shell configurations
satisfying (\ref{on-shell-condition}) do not vanish
so that if there is a difference in the term
it cannot be redefined away.
Therefore the ${\rm Tr} B^2 F^2$ terms
in (\ref{L-hat-t}) and (\ref{(B+F)^4-t})
must coincide, which determines the value of $t$.
The result is
\begin{equation}
t=1.
\end{equation}

\section{Determination of the $O(\alpha'^2)$ part of $G_s$}
\setcounter{equation}{0}
\indent

As we discussed in the last part of Section 3,
the consideration at order $\alpha'^2$ is not sufficient
to determine the $O(\alpha'^2)$ part of $G_s$
and that of the field redefinition which relates $\hat{A}_i$
to $A_i$ uniquely but allows the following ambiguity:
\begin{eqnarray}
G_s
&=& g_s \left[
1 +\frac{c-1}{4} (2 \pi \alpha')^2
{\rm Tr} B^2 + O(\alpha'^4)
\right],
\label{G_s-c}
\\
\hat{A}_i &=& A_i + \frac{1}{2} (2 \pi \alpha')^2
B_{kl} A_k (\partial_l A_i + F_{li})
+ \frac{c}{8} (2 \pi \alpha')^2 {\rm Tr} B^2 A_i
+ O(\alpha'^4),
\label{A-hat-c}
\end{eqnarray}
where $c$ is an undetermined constant.
In this appendix, we determine the value of $c$
by the consideration at order $\alpha'^4$.

We should first note that whatever ordering
of the field strengths we choose,
the $\ast$ product between the field strengths
in the $\hat{F}^4$ terms of (\ref{L-hat-4})
does not affect $O(\alpha'^4)$ terms, namely,
\begin{eqnarray}
(2 \pi \alpha')^2 {\rm Tr} ( G^{-1} \hat{F} )^4_{\rm arbitrary}
&=& (2 \pi \alpha')^2 {\rm Tr} ( G^{-1} \hat{F} G^{-1} \hat{F}
G^{-1} \hat{F} G^{-1} \hat{F} )
+ O(\alpha'^6), \\
(2 \pi \alpha')^2
( {\rm Tr} ( G^{-1} \hat{F} )^2 )^2_{\rm arbitrary}
&=& (2 \pi \alpha')^2 [
{\rm Tr} ( G^{-1} \hat{F} G^{-1} \hat{F} ) ]^2
+ O(\alpha'^6),
\end{eqnarray}
where the product between the field strengths
on the right-hand sides of these expressions
is the ordinary one, not the $\ast$ product.
Now the $\hat{F}^4$ terms in (\ref{L-hat-4})
are evaluated in the $\alpha'$ expansion
using (\ref{G-expansion}), (\ref{star-expansion}),
(\ref{G_s-c}) and (\ref{A-hat-c})
as follows:
\begin{eqnarray}
&& \frac{\sqrt{\det G}}{G_s} 
\biggl[
\frac{1}{2} (2 \pi \alpha')^2
{\rm Tr} ( G^{-1} \hat{F} )^4_{\rm arbitrary}
-\frac{1}{8} (2 \pi \alpha')^2
( {\rm Tr} ( G^{-1} \hat{F} )^2 )^2_{\rm arbitrary}
\biggr]
\nonumber \\
&=& \frac{\sqrt{\det g}}{g_s} 
\Biggl[
\frac{1}{2} (2 \pi \alpha')^2  {\rm Tr} F^4
-\frac{1}{8} (2 \pi \alpha')^2 ( {\rm Tr} F^2 )^2
\nonumber \\
&& + (2 \pi \alpha')^4 \biggl[
2 {\rm Tr} B F^5 -\frac{1}{4} {\rm Tr} BF {\rm Tr} F^4
-\frac{1}{2} {\rm Tr} BF^3 {\rm Tr} F^2
+\frac{1}{16} {\rm Tr} BF ( {\rm Tr} F^2 )^2
\biggr]
\nonumber \\
&& + (2 \pi \alpha')^4 \biggl[
2 {\rm Tr} B^2 F^4
+ \frac{c-1}{8} {\rm Tr} B^2 {\rm Tr} F^4
-\frac{1}{2} {\rm Tr} B^2 F^2 {\rm Tr} F^2
+\frac{1-c}{32} {\rm Tr} B^2 ( {\rm Tr} F^2 )^2
\biggr]
\nonumber \\
&& + {\rm ~total~derivative} + O(\alpha'^6) \Biggr].
\label{L-hat-6}
\end{eqnarray}
Since no other terms can produce
$O(B F^5)$ terms,
(\ref{L-hat-6}) gives the complete form
of them on the non-commutative side.
On the other hand, there are three sources for $O(BF^5)$ terms
on the commutative side, which are
\begin{eqnarray}
{\rm Tr} (B+F)^6 &=& {\rm Tr} F^6 + 6 {\rm Tr} B F^5 + O(B^2), \\
{\rm Tr} (B+F)^2 {\rm Tr} (B+F)^4
&=& {\rm Tr} F^2 {\rm Tr} F^4
\nonumber \\
&& + 2 {\rm Tr} B F {\rm Tr} F^4
+ 4 {\rm Tr} F^2 {\rm Tr} B F^3 + O(B^2), \\
\left[ {\rm Tr} (B+F)^2 \right]^3
&=& ( {\rm Tr} F^2 )^3 + 6 {\rm Tr} B F ( {\rm Tr} F^2 )^2
+ O(B^2).
\end{eqnarray}
By comparison, we can see that
the $O(BF^5)$ terms in (\ref{L-hat-6}) are reproduced by
the following terms in ${\cal L}(B+F)$:
\begin{equation}
\frac{(2 \pi \alpha')^4 \sqrt{\det g}}{g_s} \left[
\frac{1}{3} {\rm Tr} (B+F)^6
- \frac{1}{8} {\rm Tr} (B+F)^2 {\rm Tr} (B+F)^4
+ \frac{1}{96} \left[ {\rm Tr} (B+F)^2 \right]^3
\right].
\label{(B+F)^6}
\end{equation}
These are precisely the terms needed to take the form of
the DBI Lagrangian (\ref{DBI}) under our normalization convention.
To show that this is the unique structure of the $F^6$ terms
consistent with the assumptions (\ref{two}),
we must verify that no solution to
the on-shell initial term condition (\ref{on-shell-initial})
is possible in the $F^6$ terms.
However, even if there exist such solutions,
although we believe that there is none,
the resulting ambiguity does not affect
the determination of the $O(\alpha'^2)$ part of $G_s$
since solutions to the on-shell initial term condition
by definition
do not contribute to the $B$-dependent part of the S-matrix.
Thus the argument which has been made so far is
sufficient for the determination.

Now let us compare the $O(B^2)$ part of (\ref{(B+F)^6}),
\begin{eqnarray}
&& \frac{(2 \pi \alpha')^4 \sqrt{\det g}}{g_s} \left[
2 {\rm Tr} B^2 F^4
- \frac{1}{8} {\rm Tr} B^2 {\rm Tr} F^4
- \frac{1}{2} {\rm Tr} F^2 {\rm Tr} B^2 F^2
+ \frac{1}{32} {\rm Tr} B^2 ( {\rm Tr} F^2 )^2
\right. \nonumber \\ && \qquad \qquad \qquad \qquad
+ 2 {\rm Tr} B F B F^3
+ {\rm Tr}  B F^2 B F^2
- {\rm Tr} B F {\rm Tr} B F^3
\nonumber \\ && \qquad \qquad \qquad \qquad \left.
- \frac{1}{4} {\rm Tr} F^2 {\rm Tr} B F B F
+ \frac{1}{8} {\rm Tr} F^2 ( {\rm Tr} B F )^2
\right],
\label{B^2-in-(B+F)^6}
\end{eqnarray}
with the corresponding one on the non-commutative side.
In addition to the $O(B^2)$ part of (\ref{L-hat-6})
at order $\alpha'^4$,
there is the following contribution from the $\hat{F}^2$ term:
\begin{eqnarray}
\frac{(2 \pi \alpha')^4 \sqrt{\det g}}{g_s}
\left[
{\rm Tr} B F^2 B F^2 + 2 A_k B_{kl} \partial_l F_{ij} (FBF)_{ji}
+ A_k B_{kl} \partial_l F_{ij} A_n B_{nm} \partial_m F_{ji}
\right. \nonumber \\ \left.
+ 2 B_{kl} B_{nm} A_n ( \partial_m A_i + F_{mi} )
\partial_l A_j \partial_k F_{ji}
+ O(B^3) + {\rm total~derivative}
\right].
\end{eqnarray}
By the comparison with respect to
the terms ${\rm Tr} B^2 {\rm Tr} F^4$
and ${\rm Tr} B^2 ( {\rm Tr} F^2 )^2$
which obviously contribute to the S-matrix,
the value of the constant $c$ is determined.
The result is
\begin{equation}
c=0.
\end{equation}
To complete the argument of the $O(B^2)$ part
at order $\alpha'^4$,
it is necessary to verify that the difference,
\begin{eqnarray}
&& \frac{(2 \pi \alpha')^4 \sqrt{\det g}}{g_s} \biggl[
2 {\rm Tr} B F B F^3
- {\rm Tr} B F {\rm Tr} B F^3
- \frac{1}{4} {\rm Tr} F^2 {\rm Tr} B F B F
+ \frac{1}{8} {\rm Tr} F^2 ( {\rm Tr} B F )^2
\nonumber \\
&& \qquad \qquad \qquad \quad
- 2 A_k B_{kl} \partial_l F_{ij} (FBF)_{ji}
- A_k B_{kl} \partial_l F_{ij} A_n B_{nm} \partial_m F_{ji}
\nonumber \\
&& \qquad \qquad \qquad \quad
- 2 B_{kl} B_{nm} A_n ( \partial_m A_i + F_{mi} )
\partial_l A_j \partial_k F_{ji}
\biggr],
\label{difference}
\end{eqnarray}
does not contribute to the S-matrix and is absorbed
by a field redefinition of $\hat{A}_i$ at order $\alpha'^4$.
This is an interesting problem itself
since it is related to the $O(\theta^2)$ part
of (\ref{A-hat-2}).
However it is easily seen that
it is irrelevant to the determination
of the constant $c$ at any rate
because none of the terms in (\ref{difference})
have the structure of the contraction
${\rm Tr} B^2 = B_{ij} B_{ji}$ which was relevant
to the determination.


\end{document}